\documentclass[fleqn]{article}
\usepackage{amssymb,latexsym}
\usepackage{amsmath, amsthm}
\usepackage[dvips]{graphicx}
\DeclareFontFamily{OT1}{rsfs}{} \DeclareFontShape{OT1}{rsfs}{m}{n}{
<-7> rsfs5 <7-10> rsfs7 <10-> rsfs10}{}
\DeclareMathAlphabet{\mycal}{OT1}{rsfs}{m}{n}
\begin{document}
\title{\bf CYK Tensors, Maxwell Field
   and Conserved Quantities for Spin-2 Field}
\author{Jacek Jezierski\thanks{Partially supported by
 Swiss National Fonds.
 E-mail: Jacek.Jezierski@fuw.edu.pl}\\
Department of Mathematical Methods in Physics, \\ University of Warsaw,
ul. Ho\.za 74, 00-682 Warsaw, Poland}
\date{PACS numbers: 11.10.Ef, 04.20.Ha, 11.30.Jj}
\maketitle


{\catcode `\@=11 \global\let\AddToReset=\@addtoreset}
\AddToReset{equation}{section}
\renewcommand{\theequation}{\thesection.\arabic{equation}}

\newtheorem{Definition}{Definition}
\newtheorem{Lemma}{Lemma}
\newtheorem{Theorem}{Theorem}

\newcommand{\hst}{{\breve{h}}} 
\newcommand{\dtwo}{\mathbf{\Delta}}
\newcommand{\kolo}[1]{\vphantom{#1}\stackrel{\circ}{#1}\!\vphantom{#1}}
\newcommand{\eq}[1]{(\ref{#1})}
\def\scri{{\mycal I}}
\def\scrip{\scri^{+}}%
\newcommand{\TBR}{T^{\scriptscriptstyle BR}}
\newcommand{\TEM}{T^{\scriptscriptstyle EM}}
\newcommand{\QBR}{CQ^{\scriptscriptstyle BR}}
\newcommand{\QEM}{CQ^{\scriptscriptstyle EM}}
\newcommand{\QYK}{\Theta}
\newcommand{\rd}{{\rm d}} 
\newcommand{\E}[1]{{\rm e}^{#1}}
\newcommand{\base}[2]{{{\partial}\over {\partial{#1}^{#2}}}}
\newcommand{\be}{\begin{equation}}
\newcommand{\ee}{\end{equation}}
\newcommand{\ber}{\begin{eqnarray}}
\newcommand{\eer}{\end{eqnarray}}
\begin{abstract}
Starting from
an important application of Conformal Yano--Killing tensors
for the existence of global
charges in gravity (which has been performed in \cite{JJspin2}
and \cite{kerrnut}), some new observations at $\scri^+$ are given.
They allow to define asymptotic charges (at future null infinity)
in terms of the Weyl tensor together with their fluxes
through $\scri^+$. It occurs that some of them play a role
of obstructions for the existence of angular momentum.
 Moreover,
new relations between solutions of the Maxwell equations
and the spin-2 field are given. They are used in the
construction of new conserved quantities which are
quadratic in terms of the Weyl tensor. The obtained
formulae are similar to the functionals obtained from the
 Bel--Robinson tensor.
\end{abstract}

\section{Introduction}

The global charges result in a natural way from a geometric formulation of
the ``Gauss law'' for the gravitational charges, defined in terms of
the Riemann tensor. They lead
to the notion of the {\em Conformal Yano--Killing tensor}
(see \cite{JJspin2} and \cite{kerrnut}).  A Conformal
Yano--Killing (CYK) equation (\ref{CYK1}) possesses twenty-dimensional
space of solutions for flat Minkowski metric in four-dimensional
spacetime.  There is no obvious correspondence between
ten-dimensional asymptotic Poincar\'e group and the
twenty-dimensional space of CYK tensors. Only half of them (the
four-momentum vector $p_\mu$ and the angular momentum tensor
$j_{\mu\nu}$) are Poincar\'e generators.  This situation is analogous
to that of electrodynamics, where, in topologically nontrivial
regions, we have two charges (electric + magnetic) despite the fact
that the gauge group is only one-dimensional.
Let us notice, that for $n=2$ ($n$ is the dimension of spacetime) the
space of solutions of the equation (\ref{CYK1}) is infinite and for
$n=3$ the corresponding space is only four-dimensional. Possible
dimensions are summarized below:\\[2ex]
\indent
\begin{tabular}{|r|c|c|c|}
\hline
dimension of spacetime & $n=2$ & $n=3$ & $n=4$ \\
dimension of (pseudo)euclidean group & 3 & 6 & 10 \\
dimension of conformal group & $\infty$ & 10 & 15 \\
dimension of space of CYK tensors & $\infty$ & 4 & 20 \\ \hline
\end{tabular} \\[2ex]
The above table shows that there is no obvious relation
between CYK tensors and the group.
 In the most interesting case $n=4$ CYK tensors are related to
 eleven-dimensional group of Poincar\'e transformations
enlarged by dilatation (pseudo-si\-mi\-la\-ri\-ty transformations).
Eleven-dimensional algebra of this group
allows us to construct (via the wedge product) all CYK tensors in
Minkowski spacetime.

A natural application of the CYK tensors to the description of
asymptotically flat spacetimes was proposed in \cite{JJspin2}. It
allows us to
define an asymptotic charge at spatial infinity without supertranslation
ambiguities. The existence or nonexistence of the
corresponding asymptotic CYK tensors can be  chosen as a criterion for
the classification of asymptotically flat spacetimes at spatial infinity.
 A definition of a {\em strong} asymptotic flatness, which
was presented in \cite{JJspin2}, is strongly related
to the notion of {\em asymptotic} CYK tensor.
According to this definition, a spacetime is
asymptotically flat if it admits maximal (i.e. 14-dimensional)
space of asymptotic solutions of CYK equations.
The {\em asymptotic conformal Yano--Killing tensor} introduced in
\cite{JJspin2} was analyzed for the Schwarz\-schild metric
in \cite{kerrnut}, and it was
 shown that this metric (and other
metrics which are asymptotically ``Schwarzschildean'' up to $O(1/r^2)$ at
spatial infinity) is among the metrics fulfilling {\em strong} asymptotic
conditions.
It is also clear from the result that 14 asymptotic gravitational charges
are well defined on the ``Schwarzschildean'' background.
On the other hand, the concept of the  asymptotic
CYK tensor which defines conserved quantity at null infinity
is only possible for stationary spacetimes. Moreover, the
{\em news function}
is an obstruction for the existence of a conserved
quantity associated with asymptotic CYK tensor.
However, we show in Section \ref{atscri} that
one can use CYK tensors from Minkowski background metric
 and define Bondi four-momentum
together with its flux through $\scrip$ in terms
of the Weyl tensor. It is also shown that  the
same construction for angular momentum is possible if
we assume that the ``ofam'' charges are vanishing
(see table in Section \ref{atscri}).

We allowed us to remind the reader previous results
(in the new extended and improved form) about
CYK tensors to be able to compare them with our new results
which we present in this paper.
Here we show
the relations between solutions of the spin-2 field
  equations
and Maxwell fields, and apply these relations for the
construction of new conserved quantities which are
quadratic in terms of the Weyl tensor. The obtained
formulae are similar to the functionals obtained
with the help of the Bel--Robinson tensor and they should be useful
in Christodoulou--Klainerman \cite{Ch-Kl1}
 method to control asymptotic behaviour of the initial data in General
 Relativity.

 It also occurs that our new (quadratic in terms of the Weyl tensor)
 functionals can be
 nicely described by a universal object --- a new tensor with six indices.
 We also examine its formal properties. Our proposition is not
 included in the several generalizations of the Bel--Robinson tensor
 which are called superenergy tensors (see e.g. \cite{Sen}).
 One can say that we propose a new super-tensor with
 interesting properties.

 This paper is organized as follows: In the next Section we remind
 some basic facts about spin-2 field.
Section 3 contains small review about CYK tensors and
their applications to the asymptotic charges.
In Section 4 we show the relation between spin-2 field and Maxwell field.
Section 5 is devoted to the new conserved quantities which are quadratic
in terms of the spin-2 field.
To clarify the exposition some of the technical results and proofs have been
shifted to the appendix. Moreover, in Appendix A we have added, for
completeness, some general properties of CYK tensors.
We include also the list of symbols in Appendix C.

\section{Spin-2 field}
Let us start with
 the standard formulation of a spin-2 field $W_{\mu \alpha
\nu \beta}$ in Minkowski spacetime equipped with a flat
metric $\eta_{\mu\nu}$ and its inverse $\eta^{\mu\nu}$.
The field $W$ can be also
interpreted as a Weyl tensor for linearized
gravity (see \cite{Ch-Kl}, \cite{JJspin2}, \cite{JJnullweyl}).

\begin{Definition}
The following properties:
\be\label{s2W} W_{\mu \alpha \nu \beta}=W_{\nu \beta \mu \alpha} =
W_{[\mu \alpha ] [ \nu \beta ]}
\; , \; \;  W_{\mu [\alpha \nu \beta ]}=0 \; , \; \;
\eta^{\mu\nu}W{_{\mu\alpha \nu \beta}} =0 \ee
can be used as a definition of spin-2 field $W$.
\end{Definition}

The $*$--operation defined as
\[ ({^*} W)_{\alpha\beta\gamma\delta}=\frac 12 \varepsilon_{\alpha \beta
\mu \nu} W^{\mu\nu}{_{\gamma\delta}} \, , \quad
 (W^*)_{\alpha\beta\gamma\delta}=\frac 12 W_{\alpha\beta}{^{\mu\nu}}
 \varepsilon_{\mu \nu \gamma\delta}  \]
has the following properties:
\[ ({^*} W^*)_{\alpha\beta\gamma\delta}=\frac 14 \varepsilon_{\alpha \beta
\mu \nu} W^{\mu\nu\rho\sigma}\varepsilon_{\rho\sigma\gamma\delta} \, , \quad
 {^*} W =W^* \; , \; \; {^*} ({^*} W) = {^*}W^* =-W \, ,\]
where $\varepsilon_{\mu \nu \gamma\delta}$ is a Levi--Civita
skew-symmetric tensor
 and ${^*}W$ is called dual spin-2 field.
The above formulae are also valid for general Lorentzian metrics.

Moreover,
Bianchi identities play a role of field equations and we have the following
\begin{Lemma}
Field equations
\be\label{divW}
 \nabla_{[\lambda} W_{\mu\nu ] \alpha\beta} =0  \ee
are equivalent to
\[ \nabla^\mu W_{\mu\nu\alpha\beta} =0 \;\;
\mbox{or} \;\;
 \nabla_{[\lambda} {^*}W_{\mu\nu ] \alpha\beta} =0 \;\;
\mbox{or} \;\; \nabla^\mu \, {^*}W_{\mu\nu\alpha\beta} =0  \; . \]
\end{Lemma}
The equations in the above Lemma are also valid for
any Ricci flat metric and its Weyl tensor.


\section{Conformal Yano--Killing tensors}

Let $Q_{\mu\nu}$ be a skew-symmetric tensor field. Contracting the  Weyl tensor
$W^{\mu\nu\kappa\lambda}$ with $Q_{\mu\nu}$ we obtain a natural object
which can be integrated over two-surfaces. The result does not depend on the
choice of the surface if $Q_{\mu\nu}$ fulfills the following condition
introduced by Penrose
(see \cite{Pen-Rin} and \cite{JNG}):
\begin{equation}\label{Q0}
Q_{\lambda (\kappa ;\sigma)} -Q_{\kappa (\lambda ;\sigma)} +
\eta_{\sigma[\lambda} Q_{\kappa ]}{^\delta}_{;\delta} =0 \, .
\end{equation}
Following \cite{JJspin2}
one can rewrite equation (\ref{Q0}) in a generalized form for
$n$-dimensional spacetime with metric $g_{\mu\nu}$:
\begin{equation}\label{Qn}
Q_{\lambda (\kappa ;\sigma)} -Q_{\kappa (\lambda ;\sigma)} +
\frac 3{n-1} g_{\sigma[\lambda} Q_{\kappa ]}{^\delta}_{;\delta} =0
\end{equation}
or in the equivalent form:
\begin{equation}\label{CYK1}
Q_{\lambda \kappa ;\sigma} +Q_{\sigma \kappa ;\lambda} =
\frac{2}{n-1} \left( g_{\sigma \lambda}Q^{\nu}{_{\kappa ;\nu}} +
g_{\kappa (\lambda } Q_{\sigma)}{^{\mu}}{_{ ;\mu}} \right) \, .
\end{equation}
Let us define
\be\label{Q3i}
{\cal Q}_{\lambda \kappa \sigma}(Q,g):=
Q_{\lambda \kappa ;\sigma} +Q_{\sigma \kappa ;\lambda} -
\frac{2}{n-1} \left( g_{\sigma \lambda}Q^{\nu}{_{\kappa ;\nu}} +
g_{\kappa (\lambda } Q_{\sigma)}{^{\mu}}{_{ ;\mu}} \right)
\ee
\begin{Definition}\label{dCYK}
A skew-symmetric tensor $Q_{\mu\nu}$
 is  a {\em conformal Yano--Killing tensor} (or simply CYK tensor)
 for the metric $g$
 iff ${\cal Q}_{\lambda \kappa \sigma}(Q,g)=0$.
\end{Definition}

 The CYK tensor is a natural
generalization of the Yano tensor\footnote{A Yano
tensor (often called Killing-Yano) fulfills stronger equation
$Q_{\lambda \kappa ;\sigma} +Q_{\sigma \kappa ;\lambda} =0$
(i.e. \eq{CYK1} with vanishing right-hand side), hence every Yano tensor
is a CYK tensor but not vice versa.
The classification of Killing-Yano tensors is given by Dietz and
Rudiger in
\cite{Dietz-Rudiger} where they also consider canonical line elements
of the metrics admitting KY tensors (cf. \cite{Collinson},
 \cite{Ibohal}). A similar problems for CYK tensors seems to be
not yet solved. Several interesting applications
of Killing-Yano tensors are proposed in \cite{Baleanu}, \cite{Rietdijk},
\cite{vanHolten},  \cite{Visinescu}.
One can also construct a scalar potentials for the Maxwell
and massless Dirac equations by using CYK tensors (see \cite{BCK}).}
 (see \cite{BF}, \cite{GR} and
\cite{Yano}) with respect to the conformal rescalings.
More precisely,
for any positive scalar function $\Omega >0$ and
for a given metric $g_{\mu\nu}$
we obtain:
\be\label{QOmega} {\cal Q}_{\lambda \kappa \sigma} (Q,g) = \Omega^{-3}
    {\cal Q}_{\lambda \kappa \sigma} (\Omega^3 Q,\Omega^2 g) \, . \ee
The formula \eq{QOmega} and the above definition
of CYK tensor gives the following
\begin{Theorem}\label{otQ}
If $Q_{\mu\nu}$ is a CYK tensor for the metric $g_{\mu\nu}$
than $\Omega^3 Q_{\mu\nu}$ is a CYK tensor for
 the conformally rescaled metric $\Omega^2 g_{\mu\nu}$.
\end{Theorem}

 It is interesting to notice, that a tensor
 $A_{\mu\nu}$ --- a ``square'' of the CYK tensor $Q_{\mu\nu}$
 defined as follows:
\[ A_{\mu\nu} := Q_{\mu}{^\lambda}Q_{\lambda \nu} \] \noindent
fulfills the following equation:
\begin{equation}\label{Kt}
A_{(\mu\nu;\kappa)} = g_{(\mu\nu} A_{\kappa)}
\quad \mbox{with} \quad A_{\kappa}=
 \frac 2{n-1} Q_{\kappa}{^\lambda}
Q_\lambda{^\delta}{_{ ; \delta}}
\end{equation}
which simply means that the symmetric tensor $A_{\mu\nu}$ is a conformal
Killing tensor. This can be also described by the following
\begin{Theorem}
If $Q_{\mu\nu}$ is a skew-symmetric conformal Yano--Killing tensor than
$A_{\mu\nu} := Q_{\mu}{^\lambda}Q_{\lambda \nu}$ is
a symmetric conformal Killing tensor.
\end{Theorem}

\underline{Remark} We show at the end of Appendix \ref{AA} that
CYK tensor is a solution of the following conformally invariant
equation:
\[ (\Box +\frac16 R)Q=\frac12 W (Q,\cdot)  \, . \]
Moreover, we show in the same Appendix that if $Q$ is a CYK
tensor and the metric is Ricci flat then
$K^\mu:=Q^{\mu\lambda}{_{;\lambda}}$ is a Killing vector field.

For our purposes
we need to specify the formulae (\ref{Qn}) and (\ref{CYK1})
to the
special case of the flat four-dimensional Minkowski space
($g_{\mu\nu}=\eta_{\mu\nu}$, $n=4$). In this simple situation the general
CYK tensor  assumes the following form in Cartesian
coordinates $(x^\mu )$:
 \begin{equation}\label{Qs}
Q^{\mu\nu} = q^{\mu\nu} + 2u^{[ \mu}x^{\nu ]} -\varepsilon^{\mu\nu}
{_{\kappa\lambda}} v^{\kappa}x^{\lambda} -\frac 12 k^{\mu\nu} x_{\lambda}
x^{\lambda} +2k^{\lambda [\nu} x^{\mu ]} x_{\lambda} \, ,
\end{equation}
where $q^{\mu\nu}$, $k^{\mu\nu}$ are constant skew-symmetric tensors and
$u^{\mu}$, $v^\mu$ are constant vectors.

It is easy to verify that the charge given by $Q_{\mu\nu}$ is well
defined. Indeed, we have:
\[ \int_{\partial V} W^{\mu\nu\lambda\kappa}Q_{\lambda\kappa}
 {\rm d}S_{\mu\nu} =
\int_{V}
( W^{\mu\nu\lambda\kappa}Q_{\lambda\kappa}),_{\nu}
{\rm d}\Sigma_{\mu} = \]
\[ =
\int_{V} ( W^{\mu\nu\lambda\kappa},_{\nu} Q_{\lambda\kappa}+
W^{\mu\nu\lambda\kappa}Q_{\lambda\kappa},_{\nu} )
{\rm d}\Sigma_{\mu} = 0 \, ,\] \noindent
where the first term vanishes because of the field equations and
the second term vanishes because of the symmetries of the Weyl
tensor and because of equation
(\ref{Q0}). The above equality implies that the flux of the quantity
$W^{\mu\nu\lambda\kappa}Q_{\lambda\kappa}$ through any two closed
two-surfaces $S_1$ and $S_2$ is the same if there is a three-volume $V$
between them (i.e. if $\partial V = S_1 \cup S_2$). We define the charge
corresponding to the specific CYK tensor $Q$ as the value of this flux.

For any skew-symmetric tensor $t_{\lambda \kappa}$
let us define its dual $t^*_{\mu\nu}$ as follows:
\[ t^*_{\mu\nu} := \frac 12 \varepsilon_{\mu\nu}{^{\lambda\kappa}}
t_{\lambda \kappa}\, . \]
The above construction applied to the dual spin-2 field
${^*\!}W$
\[ \int_{S} {^*\!}W^{\mu\nu\lambda\kappa}Q_{\lambda\kappa}
 {\rm d}S_{\mu\nu}
  =  \int_{S} W^{\mu\nu\lambda\kappa}Q{^*}_{\lambda\kappa}
 {\rm d}S_{\mu\nu} \]
(cf. \eq{F*F}) does not give more charges
because the dual tensor $Q^*$ has the same form (\ref{Qs})
with the following  interchange:
\[ q \longleftrightarrow q^* \quad k \longleftrightarrow k^*
\quad u \longleftrightarrow v  \, . \]
Although, for generic metric, CYK equation is not invariant
with respect to the $*$--operation  the space of
solutions in Minkowski spacetime is closed with respect to Hodge dual.
We can summarize this property by the following
\begin{Lemma}\label{QQ*}
 If $Q$ is a CYK tensor in Minkowski spacetime
  than the dual tensor $Q^*$ possesses also
 a CYK property.
\end{Lemma}
Let us also observe that the solutions of equation (\ref{Qs}) form a
twenty-dimensional vector space. This means
that
\begin{Lemma}
A dimension of the space of CYK tensors in Minkowski spacetime
is 20.
\end{Lemma}
\noindent \underline{Remark} The Theorem~\ref{otQ} implies that
the above Lemma is also true for any conformally flat metric.

Let $\cal D$ be a generator of dilatations in Minkowski spacetime.
The generators
\be\label{strf}
 {\cal T}_{\mu}:=\base x\mu \; , \; \; \; {\cal L}_{\mu\nu}:=x_{\mu} \base
x\nu - x_\nu \base x\mu \; , \; \; \; {\cal D}:=x^\nu \base x\nu \ee
of pseudo-similarity group
(Poincar\'e group extended by scaling transformation)
obey the following commutation relations:
\[ [{\cal T}_\mu , {\cal T}_\nu]=0 \, , \quad
 [{\cal T}_\mu, {\cal L}_{\alpha\beta}]=\eta_{\mu\alpha}{\cal T}_\beta
-\eta_{\mu\beta}{\cal T}_\alpha \, , \quad
 [{\cal T}_\mu , {\cal D}]={\cal T}_\mu \, , \quad
 [{\cal D}, {\cal L}_{\alpha\beta}] = 0 \, , \]
\[
[{\cal L}_{\mu\nu}, {\cal L}_{\alpha\beta}] =
\eta_{\mu\alpha}{\cal L}_{\beta\nu} -\eta_{\mu\beta}{\cal L}_{\alpha\nu}+
\eta_{\nu\alpha}{\cal L}_{\mu\beta}-\eta_{\nu\beta}{\cal L}_{\mu\alpha}
\, . \]
The above algebra allows us to define a natural basis in the
twenty-dimensional space
of CYK tensors in Minkowski spacetime:
${\cal T}_\mu\wedge{\cal T}_\nu$, ${\cal D}\wedge{\cal T}_\mu$,
$({\cal D}\wedge{\cal T}_\mu){^*}$,
${\cal D}\wedge{\cal L}_{\mu\nu}
-\frac 12 \eta({\cal D},{\cal D}) {\cal T}_\mu\wedge{\cal T}_\nu$
with $\mu < \nu$.

The following conserved quantities has been introduced in \cite{JJspin2}:
\be\label{wmunu}
  w_{\mu\nu}:= \frac1{16\pi}
 \int_{\partial\Sigma} W({\cal T}_\mu\wedge{\cal T}_\nu) \ee
\[  w^*_{\mu\nu} := \frac1{16\pi}
 \int_{\partial\Sigma} {^*}W({\cal T}_\mu\wedge{\cal T}_\nu) =
 \frac1{16\pi} \int_{\partial\Sigma} W^*({\cal T}_\mu\wedge{\cal T}_\nu) \]
\be\label{pmu}
 p_\mu := \frac1{16\pi}\int_{\partial\Sigma} W({\cal D}\wedge{\cal T}_\mu)
 \ee
\be\label{bmu}
 b_\mu :=\frac1{16\pi}\int_{\partial\Sigma} {^*}W({\cal D}\wedge{\cal T}_\mu)
 \ee
\be\label{jmunu}
 j_{\mu\nu} :=\frac1{16\pi}\int_{\partial\Sigma}
W\left({\cal D}\wedge{\cal L}_{\mu\nu}
-\frac 12 \eta({\cal D},{\cal D}) {\cal T}_\mu\wedge{\cal T}_\nu
\right) \ee
\[
 j^*_{\mu\nu} :=\frac1{16\pi}\int_{\partial\Sigma}
{^*}W\left({\cal D}\wedge{\cal L}_{\mu\nu}
-\frac 12 \eta({\cal D},{\cal D}) {\cal T}_\mu\wedge{\cal T}_\nu
\right) \]
\noindent The conservation law for the charge $w_{\mu\nu}$ is a consequence
of field equations:
\[ \int_{\partial\Sigma} W^{\mu\nu}{_{\lambda\kappa}}
 {\rm d}\sigma_{\mu\nu} =
\int_{\Sigma} (W^{\mu\nu}{_{\lambda\kappa}}),_{\nu}
{\rm d}\Sigma_{\mu} =0 \] \noindent
For $p^\mu$ and $b^\mu$ we obtain the conservation laws from the
following observation:
\[ \int_{\partial\Sigma} x^{\lambda}W^{\mu\nu}{_{\lambda\kappa}}
 {\rm d}\sigma_{\mu\nu} =
\int_{\Sigma} \left( x^{\lambda}W^{\mu\nu}{_{\lambda\kappa}},_{\nu} +
\delta^{\lambda}_\nu W^{\mu\nu}{_{\lambda\kappa}} \right)
{\rm d}\Sigma_{\mu} =0 \] \noindent
(the same holds for ${^*}W$).\\
For $j^{\mu\nu}$ the corresponding identities are as follows:
\[ \int_{\partial\Sigma} (x_{\lambda}W^{\mu\nu\lambda\kappa}x^{\delta}
-x_{\lambda}W^{\mu\nu\lambda\delta}x^{\kappa}
 -\frac 12 x^{\lambda}x_{\lambda} W^{\mu\nu\delta\kappa})
 {\rm d}\sigma_{\mu\nu} = \]
 \[ =\int_{\Sigma} (x_{\lambda}W^{\mu\nu\lambda\kappa}x^{\delta}
-x_{\lambda}W^{\mu\nu\lambda\delta}x^{\kappa}
 -\frac 12 x^{\lambda}x_{\lambda} W^{\mu\nu\delta\kappa} ),_{\nu}
{\rm d}\Sigma_{\mu} = \]
\[ =
\int_{\Sigma} (x_{\lambda}W^{\mu\delta\lambda\kappa}
-x_{\lambda}W^{\mu\kappa\lambda\delta} - x_{\lambda}
W^{\mu\lambda\delta\kappa} ){\rm d}\Sigma_{\mu} =  \]
\[ =\int_{\Sigma} x_{\lambda}
(W^{\mu\delta\lambda\kappa}+W^{\mu\kappa\delta\lambda} +
W^{\mu\lambda\kappa\delta}){\rm d}\Sigma_{\mu} =0 \, . \]
The charge $p^\mu$ is called energy-momentum four-vector
or shortly four-momentum, $j^{\mu\nu}$ -- angular momentum tensor and
$b^\mu$ -- dual four-momentum.
Moreover, $w^{\mu\nu}$ vanishes if we assume
that $W=O\left(\frac1{r^3}\right)$.
Hence only 14 charges remain when we pass to the
limit at spatial infinity.
However,
{\em we don't know any local argument
(i.e. using only field equations) for the vanishing
of the charge $w^{\mu\nu}$.}
We will also show in the sequel that it plays
a crucial role for the existence of angular momentum
at future null infinity.

\subsection{Asymptotic Conformal Yano--Killing tensors
 at spatial infinity}
 Let us restrict ourselves to the case
 of a Lorentzian manifold $M$ of dimension 4.
Consider an asymptotically flat spacetime at spatial infinity,
fulfilling the Einstein equations. Moreover, suppose
 that the energy-momentum tensor of the matter vanishes
around spatial infinity (``compactly supported sources'').
Let us analyze 
this situation in terms of an asymptotically flat coordinate system.
We suppose that there exists an (asymptotically
Minkowskian) coordinate system $(x^\mu)$:
\[ g_{\mu\nu} - \eta_{\mu\nu} =O( r^{-1}) \; \; \; \; \;
 g_{\mu\nu,\lambda} =O( r^{ -2}) \] \noindent
where $ r:= \left(\sum_{k=1}^{3} (x^k)^2\right)^{1/2}$.

For a general asymptotically flat metric we cannot expect that the
CYK equation
(\ref{CYK1}) admit any solution. Instead, we assume that the
tensor ${\cal Q}_{\lambda\kappa\sigma}(Q,g)$ (cf. \eq{Q3i})
has a certain asymptotic behaviour at spatial infinity
\begin{equation}\label{QAF}
 {\cal Q}_{\mu\nu\lambda} =O( r^{-c}) \, , \quad c>0 \, . \end{equation}
Following \cite{kerrnut},
 suppose that the Riemann tensor $R_{\mu\nu\kappa\lambda}$
behaves asymptotically as follows:
\[ R_{\mu\nu\kappa\lambda} =O( r^{ -2 -d})\, , \quad d>0 \, . \] \noindent
It can be easily checked (see e.g. \cite{JNG}) that the vacuum Einstein
equations imply the  equality:
 \begin{equation}\label{*R*}
 \nabla_{\lambda} \left( {^*\!}R{^*}^{\mu\lambda}{_{\alpha\beta}}
Q^{\alpha\beta} \right) =  \frac 13
R^{\mu\lambda \alpha\beta} {\cal Q}_{\alpha\beta\lambda} \, .
 \end{equation}

The left-hand side of (\ref{*R*}) defines an asymptotic global
charge provided
that the right-hand side vanishes sufficiently fast at infinity. It is easy to
check that, for this purpose, the exponents $c,d$ have to fulfill the
inequality $ c + d > 1 $.
In a typical
situation when $d=1$, the above inequality simply means that $c > 0$. In
this case a weaker condition is also possible (for example ${\cal
Q}_{\mu\nu\lambda} =O( (\ln r)^{-1-\epsilon})$ with $\epsilon >0$).
Moreover, when $Q$ is a CYK tensor
(i.e. ${\cal Q}_{\mu\nu\lambda}$ vanishes)
 the formula (\ref{*R*})
gives an exact charge (not only asymptotic):
\begin{Lemma}\label{IQ}
If $Q$ is a CYK tensor for the Ricci flat metric $g$ and $R$ is
its Riemann tensor than the integral
\[ I(Q):= \int_{S}
{^*\!}R{^*}^{\mu\lambda}{_{\alpha\beta}} Q^{\alpha\beta} \rd
S_{\mu\lambda} \]
does not depend on the choice of the closed surface $S$.
\end{Lemma}
More precisely, it gives the same number for two ``spherical''
surfaces which can be connected by three-dimensional volume
which is located in the Ricci flat region.
\begin{Definition}
 An {\em Asymptotic Conformal Yano--Killing tensor} (ACYK
tensor) for the asymptotically flat metric $g$
 is a skew-symmetric
tensor $Q_{\mu\nu}$ such that ${\cal Q}_{\mu\nu\lambda}(Q,g) \rightarrow
0$ at spatial infinity.
\end{Definition}
Suppose that  $Q_{\mu\nu}$ behaves at spatial infinity as follows
 $Q_{\mu\nu} =O( r^{a})$, 
 $Q_{\lambda \kappa , \sigma} =O( r^{a-1})$.

For constructing the ACYK tensor, we can begin with the solutions of
(\ref{Q0}) in flat Minkowski space.
The asymptotic behaviour of ACYK
tensors for the flat metric explains the following
behaviour in general case:
\[ Q_{\mu\nu} = {^{(2)}\!}Q_{\mu\nu} + {^{(1)}\!}Q_{\mu\nu} +
{^{(0)}\!}Q_{\mu\nu} \] \noindent
where ${^{(2)}\!}Q_{\mu\nu} =O( r^2)$, ${^{(1)}\!}Q_{\mu\nu} =O( r)$ and
${^{(0)}\!}Q_{\mu\nu} =O( r^{1-c})$.

It is easy to verify that $a\leq d$ is sufficient for the convergence
of the integral $I(Q)$ from Lemma \ref{IQ}. In particular, for $a=1-c$
the integral
$I({^{(0)}\!}Q)$ vanishes if we assume that $d=1$. Moreover, for $d=a=1$
any ${^{(1)}\!}Q_{\mu\nu}$ gives finite limit at spatial infinity
for the surface integral $I(Q)$ but the ACYK property is needed to be sure
that this limit does not depend on the sequence of two-dim. spheres
approaching spatial infinity.
These considerations show that the energy-momentum
four-vector and the dual one are well defined for any asymptotically
flat spacetime.

The situation becomes less trivial when we pass to the
case of  ${^{(2)}\!}Q_{\mu\nu}$ which corresponds to angular
momentum. The integral $I({^{(2)}\!}Q)$ for a generic ${^{(2)}\!}Q_{\mu\nu}$
is divergent but if ACYK property for ${^{(2)}\!}Q_{\mu\nu}$
is fulfilled it has to converge (the divergent part of the integrand
in Lemma \ref{IQ} integrates to zero). However, the existence
of nontrivial ${^{(2)}\!}Q_{\mu\nu}$ with ACYK property for any
asymptotically flat spacetime is not obvious\footnote{Let us notice
that the first derivatives of ${^{(1)}\!}Q_{\mu\nu}$
are finite (${^{(1)}\!}Q_{\mu\nu,\sigma}=O(1)$) so
it is easy to believe that for particular
${^{(1)}\!}Q_{\mu\nu}$  some combinations
(in our case ${\cal Q}_{\mu\nu\lambda}$)
of the first derivatives may have better asymptotics and they
vanish at spatial infinity. On the other hand
${^{(2)}\!}Q_{\mu\nu,\sigma}=O(r)$ and now we need to lower more
than one order which is harder.}
and this is the origin of the difficulties
with the definition of the angular momentum.

We proposed in \cite{JJspin2} a stronger definition of the asymptotic
flatness.  This definition is motivated by the above discussion.

Suppose that there exists a coordinate system $(x^\mu)$
such that:
\[ g_{\mu\nu} - \eta_{\mu\nu} =O( r^{-1}) \, , \quad
 \Gamma^{\kappa}{_{\mu\nu}} =O( r^{-2}) \, , \quad
 R_{\mu\nu\kappa\lambda} =O( r^{-3}) \, . \] \noindent
In the space of ACYK tensors fulfilling the asymptotic condition
\begin{equation}\label{Q1}
Q_{\lambda \kappa ;\sigma} +Q_{\sigma
\kappa ;\lambda} -
\frac{2}{3} \left( g_{\sigma \lambda}Q^{\nu}{_{\kappa ;\nu}} +
g_{\kappa (\lambda } Q_{\sigma)}{^{\mu}}{_{ ;\mu}} \right) =
{\cal Q}_{\lambda\kappa\sigma} =O( r^{-1})
  \end{equation}
we define the following equivalence relation:
\begin{equation}\label{rel} Q_{\mu\nu} \equiv Q_{\mu\nu}'
\Longleftrightarrow Q_{\mu\nu}  - Q_{\mu\nu}' =  O(1)
 \end{equation}
for $r \rightarrow \infty$.
We assume that the space of equivalence classes defined by (\ref{Q1}) and
(\ref{rel}) has a finite dimension $D$ as a vector space.
\begin{Definition}
A spacetime is {\em strongly} asymptotically flat at spatial infinity
iff the number $D$ of gauge equivalence classes of relation \eq{rel}
equals 14 and the total dual four-momentum $b^\mu$ vanishes.
\end{Definition}
It was shown in \cite{kerrnut} that the Schwarzschildean
metrics\footnote{e.g. Kerr metric is included in this group.
Moreover, according to \cite{ptced} and \cite{CS} there exist a
non-trivial class of metrics which are Schwarzschildean.}
 are examples of {\em strongly}
asymptotically flat spacetimes and they possess the full
set of 14 asymptotic CYK tensors.
The maximal dimension $D = 14$ corresponds to the situation where there
are no supertranslation problems
in the definition of a total angular momentum at spatial infinity.
In the case of spacetimes for which $D<14$
the lack of certain ACYK tensor means that the corresponding
asymptotic charge
is not well defined. The example of Kerr--NUT spacetime
analyzed in \cite{kerrnut} suggests that
we should also assume that the asymptotic
dual four-momentum $b^\mu$ is vanishing.

\subsection{Non-conserved charges at future null infinity}\label{atscri}
The equality \eq{*R*} rewritten in terms of the Weyl
tensor for the Einstein vacuum metric can be also used
in the radiating regime.
Let us fix the conventions related to conformal rescaling
(see e.g. \cite{HF})
at $\scri^+$.
The non-physical metric we denote by $\tilde g$
and the physical metric by $g$. They are conformally related
as follows
\[ {\tilde g}_{\mu\nu} =\Omega^2 g_{\mu\nu} \, ,
   \quad \Omega^2 {\tilde g}^{\mu\nu} = g^{\mu\nu}
   \, , \quad
 {\tilde W}^\lambda{_{\mu\nu\kappa}}({\tilde g})
={W}^\lambda{_{\mu\nu\kappa}}(g) \, , \]
where ${\tilde W}$ is the Weyl tensor for the metric $\tilde g$.
Moreover, from \eq{QOmega} we have
\[ {\cal Q}_{\lambda \kappa \sigma} (Q,g) = \Omega^{-3}
    \tilde{\cal Q}_{\lambda \kappa \sigma} (\tilde Q,\tilde g) \, ,
   \quad {\cal Q}^{\lambda \kappa \sigma} (Q,g) = \Omega^{3}
    \tilde{\cal Q}^{\lambda \kappa \sigma} (\tilde Q,\tilde g) \, ,\]
where the corresponding rescaled CYK tensors
 are defined as follows (cf. Theorem \ref{otQ})
\[ {\tilde Q}_{\alpha\beta}:= \Omega^3 Q_{\alpha\beta}
   \, , \quad {\tilde Q}^{\alpha\beta}:= \Omega^{-1} Q^{\alpha\beta} \, .
\]
From above formulae we can examine the
conformal rescalings  of the  equation
\begin{equation}\label{*W*}
 \nabla_{\lambda} \left( W^{\lambda\mu}{_{\alpha\beta}}
Q^{\alpha\beta} \right) =  \frac 13
W^{\mu\lambda \alpha\beta} {\cal Q}_{\alpha\beta\lambda}
 \end{equation}
 which is equivalent to \eq{*R*} for Einstein vacuum metric.
 Let us observe that
the right-hand side
\be\label{flux1}
 {W}^\mu{_{\lambda\alpha\beta}} {\cal Q}^{\alpha\beta\lambda}(Q,g) =
 \Omega^3
 {\tilde W}^\mu{_{\lambda\alpha\beta}}
 \tilde{\cal Q}^{\alpha\beta\lambda}
\ee
of the formula \eq{*W*} can be interpreted as the flux of the quantity
\[ I(S):= \int_{S}  W^{\mu\lambda}{_{\alpha\beta}}
Q^{\alpha\beta} \rd S_{\mu\lambda} \]
 which is defined by the
 surface integral (the left-hand side of \eq{*W*}). We assume that
 two-surface $S$ is
 a sphere and it is close to null infinity.
 We have the following transformation rules for the
 corresponding densities with respect to the conformal
 rescalings:
 \be\label{charge2}
\sqrt{-\det g}\, {W}^{\mu\lambda}{_{\alpha\beta}}
Q^{\alpha\beta}(g) = \sqrt{-\det\tilde g}\,
 \Omega^{-1}
 {\tilde W}^{\mu\lambda}{_{\alpha\beta}}
 \tilde{Q}^{\alpha\beta}(\tilde g) \, .
\ee
Moreover, the flux transforms in the similar way
\be\label{flux2}
\sqrt{-\det g}\, {W}^\mu{_{\lambda\alpha\beta}}
{\cal Q}^{\alpha\beta\lambda}(Q,g) = \sqrt{-\det\tilde g}\,
 \Omega^{-1}
 {\tilde W}^\mu{_{\lambda\alpha\beta}}
 \tilde{\cal Q}^{\alpha\beta\lambda} \, ,
\ee
and the asymptotic behaviour
\be\label{WOO}
 {\tilde W}^\mu{_{\lambda\alpha\beta}}=O(\Omega) \, \quad
  \tilde{\cal Q}^{\alpha\beta\lambda} =O(1)= \tilde{Q}^{\alpha\beta}\ee
gives the finite integrals at $\scri^+$.
The formulae (\ref{charge2}-\ref{flux2})
allow us to define the flux of the energy (or
other quantity associated with $Q$)
through the piece of  $\scri^+$ between any two cross-sections of the
null infinity. More precisely, let
$ s_i : S^2 \longrightarrow \scri^+$
 for $i=1,2$ be two different
cross-sections of $\scri^+$ such that there exists $N\subset \scri^+$ with
$\partial N=s_2(S^2) \cup s_1(S^2)$. Then we have
\[
I(s_2)-I(s_1)=\int_{\partial N}
\sqrt{-\det\tilde g}\,
 \Omega^{-1}
 {\tilde W}^{\mu\lambda}{_{\alpha\beta}}
 \tilde{Q}^{\alpha\beta}
 \partial_\mu \wedge\partial_\lambda \rfloor
 \rd y^0\wedge\ldots\wedge\rd y^3
 =\]
\be = \frac13
\int_{N}
\sqrt{-\det\tilde g}\,
 \Omega^{-1}
 {\tilde W}^\mu{_{\lambda\alpha\beta}}
 \tilde{\cal Q}^{\alpha\beta\lambda}\partial_\mu
 \rfloor
 \rd y^0\wedge\ldots\wedge\rd y^3
= (\mbox{flux through}\; N)
\label{fN} \ee

In Minkowski spacetime for $\Omega=x=\frac1r$ we have
\begin{eqnarray}
\label{eta1} \tilde\eta_{\mu\nu}\rd x^\mu \rd x^\nu
& = & -x^{2}\rd u^2 + 2\rd u \rd x + \hst_{AB}
   \rd y^A \rd y^B \\
\label{eta-1}
\tilde\eta^{\mu\nu}\partial_\mu \partial_\nu
& = & x^{2}\partial_x \partial_x + 2\partial_u \partial_x + \hst^{AB}
   \partial_A \partial_B
   \end{eqnarray}
and the corresponding CYK tensors take the following form:
$$ \begin{array}{|c|c|c|c|} \hline & & &\\[-2ex]
\mbox{charge}&\mbox{Poincar\'e}& Q & {\tilde Q} \\[1ex]
 \hline & & &\\[-2ex]
\mbox{ofam} &\mbox{---} &
{\cal T}_k\wedge {\cal T}_0 & v_{,A} \rd y^A  \wedge \rd x \\ \hline
& & &\\[-2ex]
\mbox{ofam} &\mbox{---} &
{\cal T}_k\wedge {\cal T}_l & \varepsilon_A{^B}
  {\hat v}_{,B} \rd y^A \wedge \rd x \\ \hline
& & &\\[-2ex]
\mbox{energy} &\mbox{time transl.} &
{\cal D}\wedge {\cal T}_0 & \rd u \wedge \rd x \\ \hline
& & &\\[-2ex]
\mbox{linear mom.} &\mbox{space transl.} &
 {\cal D}\wedge {\cal T}_k & v \rd u \wedge \rd x \\ \hline
 & & &\\[-2ex]
\mbox{angular} &\mbox{rotation} & \displaystyle
 {\cal D}\wedge {\cal L}_{kl}-\frac12x^\mu x_\mu
  {\cal T}_k\wedge {\cal T}_l & \varepsilon_A{^B}
  {\hat v}_{,B} \rd y^A \wedge (x^{-1}\rd x +\rd u) \\
\mbox{momentum} & & &+\frac12 {\hat v}
\varepsilon_{AB} \rd y^A \wedge \rd y^B \\
 \hline & & & \\[-2ex]
\mbox{static moment} &\mbox{boost} & \displaystyle
 {\cal D}\wedge {\cal L}_{0k}-\frac12x^\mu x_\mu
  {\cal T}_0\wedge {\cal T}_k &
  v_{,A} \rd y^A \wedge (x^{-1}\rd x +\rd u)
  \\[-2ex] & & &
   \\ \hline
\end{array}
$$
where $v=\frac{x_k}r$, ${\hat v}=\varepsilon_{klm}\frac{x^m}r$
are dipole functions on a sphere $S^2$
which is parameterized by coordinates $y^A$.
The above asymptotics (at future null infinity) for CYK tensors in Minkowski
spacetime, provided for any asymptotically flat manifold,
suggest that the energy-momentum four-vector
and its density of flux \eq{flux2} is always finite ($\tilde Q =O(1)$)
but for angular momentum, in general,
we may have divergences ($\tilde Q =O(\Omega^{-1})$).
The extra condition for finiteness of angular momentum at $\scri^+$
is related with the term:
\[ x^{-1}{\tilde W}^{\mu\lambda\alpha\beta}{\tilde Q}_{\alpha\beta}
   = \left\{ \begin{array}{ll}
   x^{-2}{\tilde W}^{uxAx} v_{,A} +O(1) & \mbox{for boost} \\
   x^{-2}{\tilde W}^{uxAx}\varepsilon_A{^B}{\hat v}_{,B} +O(1)
   & \mbox{for rotation}
   \end{array}
   \right. \]
 and it becomes finite if we assume that dipole part\footnote{The
 dipole part of a vector field on $S^2$ is its orthogonal projection
 onto the six-dimensional space of conformal vector fields which is
 simultaneously a ``first'' eigenspace
 (with unit eigenvalue) for Laplace--Beltrami
 operator $\dtwo$
 (see also Appendix E in \cite{JJnullweyl}).}
\be\label{dipW} \mbox{dip}\left({\tilde W}^{uxAx}\right)=O(x^2) \, . \ee
This also means that ofam-charges are {\em \textbf{o}bstructions}$\,$
\textbf{f}or the
existence of \textbf{a}ngular \textbf{m}omentum at $\scri^+$.
The asymptotic conditions which guarantee finiteness of
the four-momentum at future null infinity are as follows:
energy ---
$\mbox{mon}\left(x^{-1}{\tilde W}^{uxux}\right)=O(1)$\footnote{By
$\mbox{mon}$ we denote
 monopole part of a scalar field on $S^2$ i.e. its orthogonal projection
 onto the one-dimensional kernel of Laplace--Beltrami
 operator $\dtwo$
 (see also Appendix E in \cite{JJnullweyl}).},
linear momentum ---
 $\mbox{dip}\left(x^{-1}{\tilde W}^{uxux}\right)=O(1)$ and
 they are fulfilled for any asymptotically flat spacetime
 satisfying \eq{WOO}. However, for
angular momentum the asymptotic condition
$\mbox{dip}\left(x^{-2}{\tilde W}^{uxAx}\right)=O(1)$
is an extra assumption which is
fulfilled for Bondi metrics (see \cite{ptcjjk} p. 91, \cite{tafel})
but it is not obvious (see also \cite{JNG}).
More precisely, for Bondi metric in the asymptotic
 form  given in Section 5.6 of \cite{ptcjjk}
we obtain
\[ x^{-1}{\tilde W}^{uxAx} = x^{-1}{\tilde g}^{xu}{\tilde g}^{xu}
{\tilde g}^{AB}{\tilde W}^u{_{uBu}} +\mbox{higher order terms} \]
and the main asymptotic term takes the following form
\[ x^{-1}{\tilde W}^{uxAx} \sim x^{-1}h^{AB}{W}^u{_{uBu}}
\sim -x^{-2}\partial_u U^A =
\frac12\partial_u \chi^{AB}{_{||B}} + O(x)\, . \]
Hence, the asymptotics of $x^{-1}{\tilde W}^{uxAx}$ at $\scri^+$
corresponds to $\frac12\partial_u \chi^{AB}{_{||B}}$ and
the dipole part of $\chi^{AB}{_{||B}}$ vanishes
because $\chi_{AB}$ is a traceless
symmetric tensor on a unit sphere
 (see also Appendix E in \cite{JJnullweyl}).

The corresponding components of the Weyl field
 in linearized gravity derived
from \eq{weyl} (presented in Appendix) are the following:
\[ 2{\tilde W}^{uxxu} = {\bf x} \, , \quad
2{^*\!}{\tilde W}^{uxxu} = {\bf y} \, , \]
\[ 2{\tilde W}^{xuxA}{_{||A}}=\partial_u {\bf x} \, , \quad
   2{\tilde W}^{xux}{_{A||B}}\varepsilon^{AB} =\partial_u {\bf y} \]
   Moreover, in linearized theory
\[ \mbox{dip}\left(\partial_u{\bf x} \right) = 0
 = \mbox{dip}\left( \partial_u{\bf y} \right) \, .
    \]
The condition \eq{dipW} means that zero from the linear case
should be replaced (in nonlinear case) by higher order asymptotics.

\section{Spin-2 field + CYK tensor $\longrightarrow$ Maxwell field}

Let us define a skew-symmetric tensor
\be\label{deF}
F_{\mu\nu}(W,Q) := W_{\mu\nu\lambda\kappa}Q^{\lambda\kappa} \, ,
\ee
where $W$ is the spin-2 field and $Q$ is the CYK tensor.
\begin{Theorem}\label{WQF}
 For any spin-2 field $W$ satisfying field equations \eq{divW} and any
 CYK tensor $Q$ in Minkowski spacetime the skew-symmetric
 tensor $F_{\mu\nu}$ (two-form $F$)
 defined by \eq{deF} fulfills vacuum Maxwell equations i.e.
 \[ \rd F =0 = \rd F{^*} \quad \iff \quad
 \nabla_{\lambda} F{^*}^{\mu\lambda}=0=\nabla_{\lambda} F^{\mu\lambda}
 \, ,\]
\end{Theorem}\noindent
where $F{^*}^{\mu\lambda}=\frac12\varepsilon^{\mu\lambda\rho\sigma}
  F_{\rho\sigma}$. \begin{proof}
This is a simple consequence of the spin-2 field equations
and the definition of CYK tensor. More precisely, from \eq{*R*}
we have
\be\label{12rM}
 0 = - \nabla_{\lambda} \left( {^*\!}W{^*}^{\mu\lambda}{_{\alpha\beta}}
Q^{\alpha\beta} \right) = \nabla_{\lambda}
\left( W^{\mu\lambda}{_{\alpha\beta}}
Q^{\alpha\beta} \right) = \nabla_{\lambda} F^{\mu\lambda}(W,Q) \, ,
\ee
so half of Maxwell equations are proved. Moreover,
\be\label{F*F}
 F{^*}^{\mu\lambda}(W,Q) = {^*\!}W^{\mu\lambda}{_{\alpha\beta}}
Q^{\alpha\beta} =  W{^*}^{\mu\lambda}{_{\alpha\beta}}
Q^{\alpha\beta} = W^{\mu\lambda}{_{\alpha\beta}}
Q{^*}^{\alpha\beta} = F^{\mu\lambda}(W,Q{^*}) \ee
and if $Q$ is a CYK tensor than $Q{^*}$ is also a CYK tensor
(Lemma \ref{QQ*})
hence from \eq{12rM} and \eq{F*F} we get
the second half of Maxwell equations:\\ \hspace*{1cm}
 $\displaystyle 0=\nabla_{\lambda} F^{\mu\lambda}(W,Q{^*})=
      \nabla_{\lambda} F{^*}^{\mu\lambda}(W,Q) \, . $
      \end{proof}
This way for each spin-2 field we can assign 20
linearly independent Maxwell fields.
Each of them may carry electric charge which is described by
(\ref{wmunu}--\ref{jmunu}). Moreover, we can define
standard energy-momentum tensor for each of them which
is obviously quadratic in terms of $F$ so it would be
also quadratic in terms of $W$.
This phenomenon is presented in the next section.

The Theorem \ref{WQF} can be generalized for Ricci flat metric,
but we have to add an assumption that $Q{^*}$ is a CYK tensor
because the Lemma \ref{QQ*} is no longer valid in this case.

\section{Conserved quantities as quadratic polynomials
 in terms of spin-2 field}

Let us start with the standard definition of energy-momentum
tensor for Maxwell field $F$:
\be\label{TEM}
 T^{\rm\scriptscriptstyle EM}_{\mu\nu}(F) := \frac12 \left( F_{\mu\sigma}F_\nu{^\sigma}
  + F^*{_{\mu\sigma}}F^*{_\nu{^\sigma}}
 \right) = F_{\mu\sigma}F_\nu{^\sigma} -\frac14 g_{\mu\nu}
 F_{\sigma\rho}F^{\sigma\rho}
\ee
The energy-momentum tensor $T^{\rm\scriptscriptstyle EM}_{\mu\nu}(F)$
is symmetric, traceless and satisfies the following positivity
condition:\\ for any non-spacelike future-directed vector fields
$X,Y$ we have
$T^{\rm\scriptscriptstyle EM}_{\mu\nu}(F)X^\mu Y^\nu \geq 0$.\\
Straightforward from the definition we get
\be\label{TF*F}
 T^{\rm\scriptscriptstyle EM}_{\mu\nu}(F) =
 T^{\rm\scriptscriptstyle EM}_{\mu\nu}(F^*) \, . \ee
Moreover, if $F$ is a Maxwell field than
\be\label{divTEM} \nabla^\mu T^{\rm\scriptscriptstyle EM}_{\mu\nu}(F)=0 \, ,
 \ee
and if $X$ is a conformal Killing vector field than
the quantity
\[ \QEM(X,\Sigma;F):=
\int_\Sigma T^{\rm\scriptscriptstyle EM}_{\mu\nu}X^\mu\rd\Sigma^\nu \]
defines a global conserved quantity
 for the spacelike hypersurface $\Sigma$ with the
end at spacelike infinity.

Let us restrict ourselves
to the spacelike hyperplanes
$\Sigma_t := \{ x\in M \; : \; x^0=t=\mbox{const} \}$.
We use the following convention for indices:
$(x^\mu)$ $\mu=0,\dots,3$ are Cartesian coordinates in Minkowski
spacetime, $x^0$ denotes temporal coordinate and
$(x^k)$ $k=1,2,3$ are coordinates on the spacelike surface $\Sigma_t$.
If the quantity $\QEM(X,\Sigma_t)$
is finite for $t=0$ than it is constant in time.
If we want to get a positive definite
integral $\QEM$, we have to restrict ourselves to the
case of non-spacelike field $X$. We can choose
 time translation ${\cal T}_0$ or
time-like conformal acceleration ${\cal K}_0$,
where
\be\label{Kcf}
 {\cal K}_\mu := -2x_\mu{\cal D} + x^\sigma x_\sigma {\cal T}_\mu \ee
is a set of four ``pure'' conformal Killing vector fields
which should be added to the eleven fields \ref{strf}
to obtain the full 15-dimensional algebra of conformal group.
This way we get
\begin{Theorem}\label{HT+HK}
There are two conserved (in time) positive definite integrals
 $\QEM( {\cal T}_0,\Sigma_t; F)$ and $\QEM( {\cal K}_0,\Sigma_t; F)$
 for the field $F$ satisfying vacuum Maxwell equations.
\end{Theorem}
\begin{proof}
This is a simple consequence of \eq{divTEM} and traceless property
of $\TEM$ which implies
$\nabla^\mu \left( T^{\rm\scriptscriptstyle EM}_{\mu\nu}X^\nu\right)=0$
for any conformal Killing vector field $X$.
\end{proof}

Following \cite{Ch-Kl},
for the Bel--Robinson tensor defined as follows
\ber\label{BRt}
T^{\scriptscriptstyle BR}_{\mu\nu\kappa\lambda} & := &
W_{\mu\rho\kappa\sigma} W_\nu{^\rho}{_\lambda}{^\sigma} +
{W^*}_{\mu\rho\kappa\sigma} {W^*}_\nu{^\rho}{_\lambda}{^\sigma} \\
& =&
 W_{\mu\rho\kappa\sigma} W_\nu{^\rho}{_\lambda}{^\sigma}
 + W_{\mu\rho\lambda\sigma} W_\nu{^\rho}{_\kappa}{^\sigma}
-\frac18 g_{\mu\nu}g_{\kappa\lambda}
W_{\alpha\beta\gamma\delta}W^{\alpha\beta\gamma\delta} \, ,
\eer
where $W$ is a spin-2 field,
one can find
a natural generalization of the Theorem \ref{HT+HK}
which is a consequence of the  properties similar to \ref{divTEM}.
More precisely, $T^{\scriptscriptstyle BR}_{\mu\nu\kappa\lambda}$ is symmetric
and traceless in all pairs of indices. Moreover,
\[ T^{\scriptscriptstyle BR}(W)=T^{\scriptscriptstyle BR}(W^*) \, , \]
and
if $W$ is a spin-2 field than
\be\label{divTBR}
\nabla^\mu T^{\rm\scriptscriptstyle BR}_{\mu\nu\lambda\kappa}(W)=0
\, .\ee
If $X,Y,Z$ are conformal Killing vector fields than
the quantity
\[ \QBR(X,Y,Z,\Sigma_t;W):=
\int_{\Sigma_t} T^{\rm\scriptscriptstyle BR}_{\mu\nu\lambda\kappa}X^\mu
Y^\nu Z^\lambda \rd\Sigma^\kappa \]
defines a global charge at time $t$.
The quantity $T^{\scriptscriptstyle BR}(X,Y,Z,T)$ is non-negative for any non-spacelike
future-directed
vector fields $X,Y,Z,T$ whenever at most two of the vector fields
are distinct.
From above properties we obtain an extension
 of the Theorem \ref{HT+HK} for the case of spin-2 field $W$:
\begin{Theorem}\label{HTTT+HKTT}
There are four conserved (in time) positive definite integrals\\
 $\QBR({\cal T}_0,{\cal T}_0, {\cal T}_0,\Sigma_t; W)$,
 $\QBR({\cal K}_0,{\cal T}_0, {\cal T}_0,\Sigma_t; W)$,
 $\QBR({\cal K}_0,{\cal K}_0, {\cal T}_0,\Sigma_t; W)$ \\ and
 $\QBR({\cal K}_0,{\cal K}_0, {\cal K}_0,\Sigma_t; W)$
 for the spin-2 field $W$ satisfying field equations \eq{divW}.
\end{Theorem}
\begin{proof}
Similarly as in the Theorem \ref{HT+HK}, from
\eq{divTBR} and traceless property of $\TBR$ we get
$\nabla^\mu \left(T^{\rm\scriptscriptstyle BR}_{\mu\nu\lambda\kappa}X^\kappa
Y^\nu Z^\lambda \right)=0$ for any conformal Killing vector fields $X,Y,Z$.
\end{proof}
Although the last integral
$\QBR({\cal K}_0,{\cal K}_0, {\cal K}_0,\Sigma_t; W)$
was not considered by Christodoulou and Klainerman, it seems to be natural
to include this quantity in the above Theorem.

We propose the following generalization
of the above considerations, namely, let us apply Maxwell field
$F$ defined by \eq{deF} into Theorem \ref{HT+HK}. Hence
 for any spin-2 field $W$ satisfying field equations
and for each CYK tensor $Q$ we obtain two conserved
positive definite quantities:
 $\QEM( {\cal T}_0,\Sigma_t; F(W,Q))$ and
 $\QEM( {\cal K}_0,\Sigma_t; F(W,Q))$.
 Because of the duality property \eq{TF*F} for
 $T^{\rm\scriptscriptstyle EM}$
 and Lemma \ref{QQ*} the number of the functionals $\QEM$
 reduces to $2\cdot 20/2=20$.
 We will show in the sequel that not all of them are
 independent and they fulfill some relations
 (cf. (\ref{K3T}-\ref{2K2T})).
 Let us denote by $\QYK$ the following functional:
 \be\label{CQYK}
  \QYK(X,V;Q):= \int_{{V\subset\Sigma}}
   T^{\rm\scriptscriptstyle EM}_{\mu\nu}
   \bigl(F(W,Q)\bigr)X^\mu\rd\Sigma^\nu
 \ee
 It seems natural to consider the following question:\\
 {\em what is the relation between four conserved quantities
 $\QBR$ from Theorem \ref{HTTT+HKTT} and our
  functionals $\QYK$?}\\
  The answer is very simple:
 \begin{Theorem}\label{BRinEM}
 The four conserved quantities $\QBR$ from Theorem \ref{HTTT+HKTT}
 are contained in our functionals $\QYK$.
 \end{Theorem}

 This problem can be easier analyzed if we pass to
 3+1-decomposition of the spin-2 field. The ten
 independent components of $W$ split into two
 3-dimensional traceless tensors: electric part
 \[ E(X,Y):= W(X,{\cal T}_0,{\cal T}_0,Y) \]
 and magnetic part
 \[ H(X,Y):= {^*\!}W(X,{\cal T}_0,{\cal T}_0,Y) \, , \]
 and the full spin-2 field $W$ expresses in terms of $E$ and $H$
 as follows:
 \begin{eqnarray}
  W_{0kl0}=E_{kl}\, , & W_{0kij}=H_{kl}\varepsilon^l{_{ij}} \, , &
  W_{klmn}= \varepsilon^i{_{kl}}\varepsilon^j{_{mn}}E_{ij} \, .
 \end{eqnarray}
Let us notice that on the surface $\Sigma_0=\{ x^0=t=0 \}$
the vector fields ${\cal K}_0$ and ${\cal T}_0$ are parallel:
\[ {\cal K}_0 (t=0)=r^2{\cal T}_0=r^2\partial_t \]
and obviously the CYK tensor
$ {\cal D}\wedge{\cal T}_0 = r\partial_r\wedge\partial_t  $
where $r$ is a radial coordinate.
The above observations give the following result for our
integrands on the surface $t=0$:
\be\label{r2TEM} \TEM\bigl({\cal K}_0,{\cal T}_0,
F(W,Q)\bigr) =
 r^2\TEM\bigl({\cal T}_0,{\cal T}_0,
F(W,Q)\bigr)
\ee
\be\label{i0} \TEM\bigl({\cal T}_0,{\cal T}_0,
F(W,{\cal T}_i\wedge{\cal T}_0)\bigr)
= \frac12 \sum_{k=1}^3\left(
 E_{ki}E^{k}{_i} + H_{ki}H^{k}{_i} \right)\ee
\be\label{D0} \TEM\bigl({\cal T}_0,{\cal T}_0,
F(W,{\cal D}\wedge{\cal T}_0)\bigr) = \frac12 r^2 \left(
 E_{kr}E^{k}{_r} + H_{kr}H^{k}{_r} \right) \ee
\be\label{Di} \TEM\bigl({\cal T}_0,{\cal T}_0,
F(W,{\cal D}\wedge{\cal T}_i)\bigr) = \frac12 r^2
v_{,A}v_{,B}\varepsilon^{AC}\varepsilon^{BD}\left(
 E_{kC}E^{k}{_D} + H_{kC}H^{k}{_D} \right) \ee
where $v=\frac{x_i}r$.
\begin{eqnarray} \label{L0i}
 \TEM\bigl({\cal T}_0,{\cal T}_0,
F(W,{\cal D}\wedge{\cal L}_{0i}
-\frac12x^\mu x_\mu{\cal T}_0\wedge{\cal T}_i
)\bigr) & = & \\ \nonumber
 & & \hspace*{-7cm}
 (vE^{k}{_r}-rv_{||A}E^{kA})(vE_{kr}-rv_{||B}E_k{^B})
 +(vH^{k}{_r}-rv_{||C}H^{kC})(vH_{kr}-rv_{||D}H_k{^D})
 \end{eqnarray}
\begin{eqnarray} \nonumber
 \TBR\bigl({\cal K}_0,{\cal K}_0,{\cal K}_0,{\cal T}_0
\bigr) & = & r^2\TBR\bigl({\cal K}_0,{\cal K}_0,{\cal T}_0,{\cal T}_0
\bigr)= r^4\TBR\bigl({\cal K}_0,{\cal T}_0,{\cal T}_0,{\cal T}_0
\bigr)\\ & =& \label{r6TBR}
 r^6 \TBR\bigl({\cal T}_0,{\cal T}_0,{\cal T}_0,{\cal T}_0
\bigr)
\end{eqnarray}
\[ \TBR\bigl({\cal T}_0,{\cal T}_0,{\cal T}_0,{\cal T}_0
\bigr) = E_{kl}E^{kl} + H_{kl}H^{kl}\]
In Appendix \ref{xyEH} we present the details how the conserved
quantities $\QBR$ are included in $\QYK$. In particular,
the above formulae integrated by parts
on each sphere $S(r)$ give the following
relations between corresponding functionals:
\ber\label{4T}
\int_{\Sigma_0}
 \TBR\bigl({\cal T}_0,{\cal T}_0,{\cal T}_0,{\cal T}_0
\bigr)\rd^3 x &= &
 2\sum_{k=1}^3 \int_{\Sigma_0} \TEM\bigl({\cal T}_0,{\cal T}_0,
F(W,{\cal T}_k\wedge{\cal T}_0)\bigr)\rd^3 x
 \eer
\begin{eqnarray} \label{K3T}
\int_{\Sigma_0}
 \TBR\bigl({\cal K}_0,{\cal T}_0,{\cal T}_0,{\cal T}_0
\bigr)\rd^3 x & = &
 2\sum_{k=1}^3 \int_{\Sigma_0} \TEM\bigl({\cal K}_0,{\cal T}_0,
F(W,{\cal T}_k\wedge{\cal T}_0)\bigr)\rd^3 x \\
& = &
\label{K3Tbis}
 2\sum_{\mu=0}^3 \int_{\Sigma_0} \TEM\bigl({\cal T}_0,{\cal T}_0,
F(W,{\cal T}_\mu\wedge{\cal D})\bigr)\rd^3 x
\end{eqnarray}
\begin{eqnarray}\label{2K2Tbis}
 \int_{\Sigma_0}
 \TBR\bigl({\cal K}_0,{\cal K}_0,{\cal T}_0,{\cal T}_0
\bigr)\rd^3 x &= &
 2\sum_{\mu=0}^3 \int_{\Sigma_0} \TEM\bigl({\cal K}_0,{\cal T}_0,
F(W,{\cal T}_\mu\wedge{\cal D})\bigr)\rd^3 x \\
 & &  \label{2K2T} \hspace*{-2.2cm}
= \; 2\sum_{i=1}^3 \int_{\Sigma_0} \TEM\bigl({\cal T}_0,{\cal T}_0,
F(W,{\cal D}\wedge{\cal L}_{0i}
-\frac12x^\mu x_\mu{\cal T}_0\wedge{\cal T}_i)\bigr)\rd^3 x
 \end{eqnarray}
\be  \label{3K1T}
\int_{\Sigma_0}
 \TBR\bigl({\cal K}_0,{\cal K}_0,{\cal K}_0,{\cal T}_0
\bigr)\rd^3 x =
2\sum_{i=1}^3 \int_{\Sigma_0} \TEM\bigl({\cal K}_0,{\cal T}_0,
F(W,{\cal D}\wedge{\cal L}_{0i}
-\frac12x^\mu x_\mu{\cal T}_0\wedge{\cal T}_i)\bigr)\rd^3 x
 \, .
 \ee
 The formulae (\ref{4T}-\ref{3K1T}) imply the
 Theorem \ref{BRinEM}.
It is also clear from (\ref{K3T}-\ref{2K2T}) that
not all twenty functionals $\QYK$ are
linearly independent. Although we have only checked the linear
dependence at $t=0$ the conservation law implies that they are
related in the same way at any time $t$ (provided they
are finite).\\[1ex]
\underline{Remark} The following problem should be examined:
{\em How many independent functionals $\QYK$ do exist for a generic
spin-2 field?}\\[1ex]
We leave this problem opened, however, we show below
(for some examples of $\QYK$ functionals)
the method which should lead to the answer to the above
question.

We should remember
 that the electric and magnetic tensors on
$\Sigma$ are not free, they are constrained by the following
equations
\be\label{divEH} E^{kl}{_{|l}}=0 \, ,\quad H^{kl}{_{|l}}=0 \ee
which are simply tangent to $\Sigma$ parts of spin-2 field
equations \eq{divW}. This means that
one should examine the integrals $\QYK$ and $\QBR$
in terms of the unconstrained
degrees of freedom which are no longer constrained.
This can be done in a systematic way, using quasi-local
variables
${\bf x}=2W\bigl({\cal T}_0,{\cal D},{\cal D},{\cal T}_0
\bigr)$ and ${\bf y}=2{^*\!}W\bigl({\cal T}_0,{\cal D},{\cal D},{\cal T}_0
\bigr)$ which are extensively used in \cite{JJschwarzl} and
\cite{JJnullweyl}. The full spin-2 field $W$ on surface
$\Sigma$ expresses in terms of Cauchy data ${\bf x}, \partial_0{\bf x},
 {\bf y}, \partial_0{\bf y}$ due to the following
 Theorem proposed in \cite{JJnullweyl}:
 \begin{Theorem}
The linearized Riemann tensor for the vacuum Einstein equations depends
quasilocally on the invariants $({\bf x},{\bf y})$ which contain the full
information about the linearized gravitational field. Moreover, the
invariants $\bf x$ and $\bf y$ fulfill usual wave equation.
\end{Theorem}

The precise form of the Weyl field is given in Appendix \ref{xyEH}.
In particular, electric part
 $E_{kl}$ expresses in terms of ${\bf x}, \partial_0{\bf y}$:
\[ r^2E_{rr}=-r^2\eta^{AB}E_{AB}=\frac12{\bf x} \, , \quad
   r^2\varepsilon^{AB}E_{rB||A}=\frac12\partial_0{\bf y} \, ,
\quad r^3 E^{rA}{_{||A}}=-\frac12\partial_r(r{\bf x}) \]
and two-dimensional traceless part
$\kolo{E}_{AB}:={E}_{AB}-\frac12\eta_{AB}\eta^{CD}E_{CD}$
is given by \eq{EABpolar} and \eq{EABaxial}.
Similarly magnetic part $H_{kl}$ depends on ${\bf y}, \partial_0{\bf x}$:
\[ r^2H_{rr}=-r^2\eta^{AB}H_{AB}=\frac12{\bf y} \, , \quad
   r^2\varepsilon^{AB}H_{rB||A}=\frac12\partial_0{\bf x} \, ,
 \quad r^3 H^{rA}{_{||A}}=-\frac12\partial_r(r{\bf y})
\]
and traceless part $\kolo{H}_{AB}$ is given by
\eq{HABpolar} and \eq{HABaxial}.
Hence all these components of the spin-2 field
inserted into $\TBR$
and integrated by parts (see Appendix \ref{xyEH}) give the following result:
\begin{eqnarray}
 \QBR_0 & := &
 \int_V \TBR\bigl({\cal K}_0,{\cal T}_0,{\cal T}_0,{\cal T}_0
\bigr)\rd^3 x \\ & = & \int_V r^2 \bigl(
E_{kl}E^{kl} + H_{kl}H^{kl}\bigr)r^2\sin\theta\rd r
 \rd\theta\rd\varphi
 \\ &= & \nonumber \frac12 \int_V \Bigl(
 (r\dot{\bf x})(-\dtwo)^{-1}(r\dot{\bf x})
 + \partial_r(r{\bf x})(-\dtwo)^{-1}\partial_r(r{\bf x}) + {\bf x}^2
 \\ & & \nonumber
 + (r\dot{\bf y})(-\dtwo)^{-1}(r\dot{\bf y})
 + \partial_r(r{\bf y})(-\dtwo)^{-1}\partial_r(r{\bf y}) + {\bf y}^2
 \Bigr)\rd r \sin\theta\rd\theta\rd\varphi + \\ & & \label{BRV}
 + \frac12 \int_V \Bigl[
   \partial_r(r^2\dot{\bf x})\dtwo^{-1}(\dtwo +2)^{-1}
   \partial_r(r^2\dot{\bf x}) + \frac12 {\bf x}^2  \\ & & \nonumber
   + \bigl(
   \partial_r [ r\partial_r(r{\bf x})] +\frac12\dtwo{\bf x}\bigr)
   \dtwo^{-1}(\dtwo +2)^{-1} \bigl(
   \partial_r [ r\partial_r(r{\bf x})] +\frac12\dtwo{\bf x}\bigr)
    \\ & & \nonumber
   + \bigl(
   \partial_r [ r\partial_r(r{\bf y})] +\frac12\dtwo{\bf y}\bigr)
   \dtwo^{-1}(\dtwo +2)^{-1} \bigl(
   \partial_r [ r\partial_r(r{\bf y})] +\frac12\dtwo{\bf y}\bigr)
   \\ & & \nonumber +
   \partial_r(r^2\dot{\bf y})\dtwo^{-1}(\dtwo +2)^{-1}
   \partial_r(r^2\dot{\bf y}) + \frac12 {\bf y}^2
   \Bigr]\rd r \sin\theta\rd\theta\rd\varphi
\end{eqnarray}
where by dot we have denoted as usual time derivative
and similarly we analyze:
\begin{eqnarray} \nonumber
\QYK_0 & := & \int_V \TEM\bigl({\cal T}_0,{\cal T}_0,
F(W,{\cal D}\wedge{\cal T}_0)\bigr)\rd^3 x  \\ \nonumber
&=& \frac12 \int_V r^2 \left(
 E_{kr}E^{kr} + H_{kr}H^{kr} \right)r^2\sin\theta\rd r
 \rd\theta\rd\varphi \\
 & =& \label{EMV} \frac14 \int_V \Bigl(
 \partial_0(r{\bf x})(-\dtwo)^{-1}\partial_0(r{\bf x})
 + \partial_r(r{\bf x})(-\dtwo)^{-1}\partial_r(r{\bf x}) + {\bf x}^2
 \\ & & \nonumber
 + \partial_0(r{\bf y})(-\dtwo)^{-1}\partial_0(r{\bf y})
 + \partial_r(r{\bf y})(-\dtwo)^{-1}\partial_r(r{\bf y}) + {\bf y}^2
 \Bigr)\rd r \sin\theta\rd\theta\rd\varphi \, .
 \end{eqnarray}
 The above reduced expressions for the close to each other
  quadratic functionals\footnote{The other expressions
   may differ also by certain power of $r$
   according to \eq{r2TEM} and \eq{r6TBR}.}
 show the differences between them, in particular,
  $\QBR_0$ contains second derivatives of the
 reduced data $\bf x, \bf y$. This is also a typical
 attribute of all functionals associated to
 the densities (\ref{i0}-\ref{L0i}) except $\QYK_0$
 built from \eq{D0}.
 In our opinion
 the functional \eq{EMV}
 seems to be the most natural one
because it contains only first (radial and time) derivative
of our quasi-local variables. Moreover, one can easily check
the conservation law for compactly supported data
straightforward from the wave equation
which is fulfilled by our quasi-local unconstrained degrees
of freedom (see Appendix \ref{xyEH}). The functional $\QYK_0$
is also very close to the energy functional proposed in
\cite{JJschwarzl} which takes the following form
in Minkowski spacetime:
\begin{eqnarray}
 8\pi{\cal H} & = & \nonumber
 \frac1{4} \int_{\Sigma}
 \Bigl[ (r\dot{\bf x})
 \dtwo ^{-1}(\dtwo +2)^{-1} (r\dot{\bf x}) + (r\dot{\bf y})
 \dtwo ^{-1}(\dtwo +2)^{-1} (r\dot{\bf y})+
\\ & &  \label{Hxy} \hphantom{\frac1{4} \int_{\Sigma}}
 + (r{\bf x})_{,3}
 \dtwo ^{-1}(\dtwo +2)^{-1} (r{\bf x})_{,3} -
  {\bf x}  (\dtwo +2)^{-1}{\bf x}  +
\\ & &  \nonumber \hphantom{\frac1{4} \int_{\Sigma}} +  (r{\bf y})_{,3}
 \dtwo ^{-1}(\dtwo +2)^{-1} (r{\bf y})_{,3} - {\bf y}
 (\dtwo +2)^{-1} {\bf y}  \Bigr]
\rd r \sin\theta\rd\theta\rd\varphi \, .
  \end{eqnarray}
The integrals \eq{EMV} and \eq{Hxy} differ by the operator
$(\dtwo +2)^{-1}$, hence for each spherical mode
(i.e. after spherical harmonics decomposition) they
 are proportional to each other.

In nonlinear case the quadratic
functionals  \eq{CQYK} should be useful in
the so called exterior initial value problems
($V=\Sigma\setminus B(0,R)$) and
they should allow
to control asymptotic behaviour of the various
components of the Weyl tensor.

\subsection{Natural (super-)tensor}
Let us consider a tensor
\be\label{T6} \hspace*{-0.7cm}
 T_{\mu\nu\alpha\beta\gamma\delta} := \frac12 \left(
W_{\mu\sigma\alpha\beta} W_\nu{^\sigma}{_{\gamma\delta}} +
W_{\mu\sigma\gamma\delta} W_\nu{^\sigma}{_{\alpha\beta}} +
{W^*}_{\mu\sigma\alpha\beta} {W^*}_\nu{^\sigma}{_{\gamma\delta}}+
{W^*}_{\mu\sigma\gamma\delta} {W^*}_\nu{^\sigma}{_{\alpha\beta}}
\right)
  \ee
which is naturally related to our new conserved quantities
by the following equality
\[ \TEM_{\mu\nu}(F(Q))=\frac12 T_{\mu\nu\alpha\beta\gamma\delta}
    Q^{\alpha\beta}Q^{\gamma\delta} \, .\] 
Tensor $T$ has the following properties:
 \be\label{pt1} T_{\mu\nu\alpha\beta\gamma\delta}=
   T_{\mu\nu\gamma\delta\alpha\beta}=
   T_{(\mu\nu)[\alpha\beta][\gamma\delta]}
    \, , \quad
 T_{\mu\nu\alpha\beta\gamma\delta}g^{\mu\nu} = 0 \ee
which are simple consequences of the definition \eq{T6} and
spin-2 field properties.
Moreover, $T$ is related with Bel--Robinson tensor
as follows
\[ g^{\beta\delta}T_{\mu\nu\alpha\beta\gamma\delta} =
   \TBR_{\mu\nu\alpha\gamma} \, . \]
One  can also show the following properties of tensor $T$:
\be\label{pt2}
 \nabla^\mu T_{\mu\nu\alpha\beta\gamma\delta} = 0
 \, , \quad T_{\mu[\nu\alpha\beta]\gamma\delta}=0 \, . \ee
\begin{proof}
The divergence-free property for $T$
is a consequence of spin-2 field equations
which simultaneously hold for $W$ and $W^*$, hence we get
\[ \nabla^\mu T_{\mu\nu\alpha\beta\gamma\delta} =
   \frac14 \nabla_\nu\left( W^{\mu\sigma}{_{\alpha\beta}}
   W_{\mu\sigma\gamma\delta} +
   {W^*}^{\mu\sigma}{_{\alpha\beta}}
   {W^*}_{\mu\sigma\gamma\delta} \right)=0 \]
 where the last equality is a consequence of
 the following formula
 \[  W^{\mu\sigma}{_{\alpha\beta}}
   W_{\mu\sigma\gamma\delta} +
   {W^*}^{\mu\sigma}{_{\alpha\beta}}
   {W^*}_{\mu\sigma\gamma\delta}  =0 \]
 which is equivalent to traceless attribute of $T$ in \eq{pt1}
 and can be easily checked from properties of spin-2 field
 with respect to $*$--operation (e.g. $*^2=-1$).
 The second equality in \eq{pt2} is implied by the Bianchi identity
 for $W$ and $W^*$.
 \end{proof} 

The above properties of tensor $T$ allow us to check
the following final
\begin{Theorem}
If $P,Q$ are CYK tensors, $X$ is a conformal vector field
and $T$ obeys the properties \eq{pt1} and \eq{pt2}
then
\[ \nabla^\mu \left( T_{\mu\nu\alpha\beta\gamma\delta}
X^\nu P^{\alpha\beta}Q^{\gamma\delta}\right) =0 \, .\]
\end{Theorem}


\section*{Acknowledgments}
The author is much indebted to P. Chru\'sciel
 for fruitful discussions and to
D\'epartement de Math\'ematiques at
Universit\'e de Tours
for the hospitality during the preparation of this paper.

\appendix
\section{General properties of conformal Yano--Killing tensors}
\label{AA}
 In this appendix, following \cite{kerrnut}, we remind some general identities
for Killing vector fields and conformal Killing vectors
together with the CYK tensors properties.

Let $M$ be a differential manifold of dimension $n>1$ with a riemannian or
 pseudoriemannian metric $g_{\mu\nu}$.
 By ``$;$'' we denote the covariant derivative associated with
the Levi--Civita connection, and by $R^{\sigma}{_{\mu\lambda\nu}}$
 we denote the curvature Riemann tensor.
$R_{\mu\nu}$ is the Ricci tensor.

If we have a tensor object $T_{\ldots\mu\nu\ldots}$ then by
 $T_{\ldots(\mu\nu)\ldots}$ we denote the symmetric part and
 by $T_{\ldots[\mu\nu]\ldots}$ the skewsymmetric part of
 $T_{\ldots\mu\nu\ldots}$ with respect to the pair of indices $\mu\nu$.
 The indices are raised and lowered with
respect to the metric $g_{\mu\nu}$ or its inverse.

\subsection{Killing vectors}
The Killing vector field $X^\mu$ on $M$ can be defined as a
 solution of the following equation:
\be\label{KV} X_{\lambda;\mu} + X_{\mu;\lambda} = 0 \, .\ee
Let us write explicitly three similar identities for any covector field
$X_\mu$:
\[ X_{\lambda;\mu\nu} - X_{\lambda;\nu\mu}
= X_\sigma R^\sigma{_{\lambda\mu\nu}} \]
\[ X_{\mu;\lambda\nu} - X_{\mu;\nu\lambda}
= X_\sigma R^\sigma{_{\mu\lambda\nu}} \]
\[ X_{\nu;\lambda\mu} - X_{\nu;\mu\lambda}
= X_\sigma R^\sigma{_{\nu\lambda\mu}} \]
From the above equalities we can derive as follows:
\[ -2X_{\mu;\nu\lambda} =
(X_{\lambda;\mu\nu}+X_{\mu;\lambda\nu})-(X_{\lambda;\nu\mu}+X_{\nu;\lambda\mu})
+(X_{\nu;\mu\lambda}+X_{\mu;\nu\lambda}) -2X_{\mu;\nu\lambda} = \]
\[ = (X_{\lambda;\mu\nu} - X_{\lambda;\nu\mu}) +(X_{\mu;\lambda\nu} -
X_{\mu;\nu\lambda})  - (X_{\nu;\lambda\mu} - X_{\nu;\mu\lambda})=\]
\[
= X_\sigma \left( R^\sigma{_{\lambda\mu\nu}} +R^\sigma{_{\mu\lambda\nu}}
 -R^\sigma{_{\nu\lambda\mu}} \right) = \]
 \[ = -X_\sigma \left( R^\sigma{_{\lambda\mu\nu}} +R^\sigma{_{\mu\nu\lambda}}
 +R^\sigma{_{\nu\lambda\mu}} \right) +2X_\sigma R^\sigma{_{\lambda\mu\nu}}=
  -2X_\sigma R^\sigma{_{\lambda\nu\mu}} \, .\]
  The above manipulations lead to the following second order equation:
 \be \label{xd2} X_{\mu;\nu\lambda}=X_\sigma R^\sigma{_{\lambda\nu\mu}}\, .
  \ee
 The trace of the above equality gives:
 \be \label{x3}
   X_{\mu;\nu\lambda}g^{\nu\lambda}=X_\sigma R^{\sigma\nu}{_{\nu\mu}}=
  -X_\sigma R^\sigma{_{\mu}} \, .
\ee
 So, if the Ricci tensor vanishes than
 any covector Killing field $X_\mu$ is a harmonic one-form:
 \[ X_\mu {^{;\nu}}{_\nu}= -X_\sigma R^\sigma{_{\mu}}=0 \, .
 \]

\subsection{Conformal Killing vectors}
A natural conformal generalization of the equation  (\ref{KV}) has the
following form:
\[ Z_{\lambda;\mu} + Z_{\mu;\lambda}
-\frac 2n g_{\lambda\mu}Z^{\sigma}{_{;\sigma}} = 0 \, .\]
The solution $Z^\lambda$ of this equation we call {\em conformal Killing
vector field}.
Let us denote $Z:=Z^{\sigma}{_{;\sigma}}$.
We can perform the similar trick as for Killing vectors, namely:
\[
Z_{\lambda;\mu\nu}+Z_{\mu;\lambda\nu}-Z_{\lambda;\nu\mu}-Z_{\nu;\lambda\mu}
+Z_{\nu;\mu\lambda}+Z_{\mu;\nu\lambda} -2Z_{\mu;\nu\lambda} = -2Z_\sigma
R^\sigma{_{\lambda\nu\mu}} \, ,\]
and we obtain the following expression for second derivatives:
\be\label{zd2} Z_{\mu;\nu\lambda} +\frac 1n \left( g_{\lambda\nu}Z_{;\mu}
 -g_{\lambda\mu}Z_{;\nu} -g_{\mu\nu}Z_{;\lambda} \right)
=Z_\sigma R^\sigma{_{\lambda\nu\mu}} \, .\ee
Taking the trace in the indices $\mu\lambda$ we get
\[ Z^\mu{_{;\nu\mu}} - Z_{;\nu}= Z_\sigma R^\sigma{_\nu}  \]
which is trivial if we remember that $Z=Z^{\sigma}{_{;\sigma}}$.

The trace with respect to the indices $\mu\nu$ in equation (\ref{zd2})
gives the identity, but the trace with respect
 to $\nu\lambda$ leads to the result:
\be Z_\mu{^{;\nu}}{_\nu}+\frac{n-2}n Z_{;\mu}= -Z_\sigma R^\sigma{_{\mu}}
\, .
\ee

If we take the derivative with respect to the index $\mu$ (a contraction)
in equation  (\ref{zd2}) and perform some further
 straightforward manipulations, we obtain the following result:
\be\label{Ztr} Z_{;\lambda\nu} +2 Z^{\sigma}{_{;(\lambda}}R_{\nu)\sigma}
+ Z^{\sigma}R_{\lambda\nu;\sigma}
 + \frac 1n \left( g_{\lambda\nu}Z^{;\mu}{_\mu} -Z_{;\nu\lambda}
-Z_{;\lambda\nu} \right) =0 \, .\ee
 We can also take a trace in (\ref{Ztr}):
\[ Z^{;\mu}{_\mu} +\frac n{2(n-1)} \left(\frac2n Z
R + Z^\sigma R_{;\sigma} \right)=0 \, ,
\]
where we used the following identity for the conformal vector field:
\[ Z R=n Z_{\mu;\nu} R^{\mu\nu} \, .\]
And finally the second derivatives of $Z$ fulfill the property:
\[ \frac{n-2}n Z_{;\lambda\nu} -\frac 1{n-1}g_{\lambda\nu}
 \left( \frac1n Z R + \frac 12 Z^\sigma R_{;\sigma} \right)
+ Z^\sigma R_{\lambda\nu;\sigma} +2 Z^{\sigma}{_{;(\lambda}}R_{\nu)\sigma}
= 0 \, .\]

\subsection{CYK tensors}

\be\label{CYKf} f_{\lambda\mu;\nu}+f_{\nu\mu;\lambda} =\frac 2{n-1} \left(
g_{\nu\lambda} f^\sigma{_{\mu;\sigma}} +
g_{\mu(\lambda}f_{\nu)}{^\sigma}{_{;\sigma}} \right) \ee
According to the Definition \ref{dCYK}
the skew-symmetric tensor $f_{\mu\nu}$ fulfilling the equation (\ref{CYKf})
  we call the {\em conformal Yano--Killing tensor}.

Let us denote $f_\kappa :=f^\sigma{_{\kappa;\sigma}}$.
By the similar trick as for (\ref{xd2}) and (\ref{zd2}) we get:
\[ 2f_{\lambda\kappa;\nu\mu}= \frac 2{n-1} \left(
g_{\lambda\mu}f_{\kappa;\nu}+g_{\nu\lambda} f_{\kappa;\mu}
- g_{\mu\nu}f_{\kappa;\lambda} -g_{\kappa(\lambda}f_{\mu);\nu}
+ g_{\kappa(\mu}f_{\nu);\lambda} -g_{\kappa(\nu}f_{\lambda);\mu}
 \right)+ \]
 \be\label{fd2} + f_{\sigma\lambda}R^\sigma{_{\kappa\mu\nu}}
 + f_{\sigma\mu}R^\sigma{_{\kappa\lambda\nu}}
 + f_{\sigma\nu}R^\sigma{_{\kappa\lambda\mu}}
 + 2 f_{\sigma\kappa}R^\sigma{_{\mu\nu\lambda}} \, .\ee
 The trace in $\lambda \nu$ in equation  (\ref{fd2})
 leads to the identity but
 the trace in $\kappa\nu$ gives the following result:
\be\label{sym1} g_{\mu\nu}f^\lambda{_{;\lambda}}+ (n-2)f_{(\mu;\nu)}=
 (n-1)R_{\sigma(\mu}f_{\nu)}{^\sigma} \, .
\ee
The trace of the  equation (\ref{sym1}) gives the equation:
 \[ f^\lambda{_{;\lambda}}= - \frac 12 f^{\mu\nu}R_{\mu\nu}=0 \, .
\]
\underline{Remark:} If $R_{\mu\nu}=0$ then from equation (\ref{sym1})
$f_{(\mu;\nu)}=0$,  so $f_{\mu}$ is a Killing field.\\[1ex]
 The trace with respect to the indices $\mu\nu$ in
equation  (\ref{fd2}) gives as follows:
\be \label{sym2} g_{\mu\nu}f^\lambda{_{;\lambda}}+ (n-2)f_{(\mu;\nu)}=
(n-1)R_{\lambda(\mu\nu)\sigma}f^{\lambda\sigma}
+(n-1)R_{\sigma(\mu}f_{\nu)}{^{\sigma}} \, ,
\ee
 \be\label{as2} f_{\lambda\kappa}{^{;\mu}{_\mu}}=
f_\sigma{^\mu}R^{\sigma}{_{[\kappa\lambda]\mu}}
+ \frac{n-4}{n-1}f_{[\lambda;\kappa]}
-f^\sigma{_{[\kappa}}R_{\lambda]\sigma} \, ,
\ee
 where we have written the symmetric and skew-symmetric parts separately.
The equations (\ref{sym1}) and (\ref{sym2}) are equivalent because
$R^{[\lambda}{_{(\mu\nu)}}{^{\sigma]}} =0$, so
$R_{\lambda(\mu\nu)\sigma}f^{\lambda\sigma}=0$.

Let us notice that for $n=4$ the second term on the right-hand side of
the equation (\ref{as2}) vanishes and the final form of this equality is
the following:
 \be\label{fwave} f_{\lambda\kappa}{^{;\mu}{_\mu}}=
\frac 12 f_{\sigma\mu}W^{\sigma\mu}{_{\lambda\kappa}}
 - \frac16 R f_{\lambda\kappa} \, ,
\ee
where by $W_{\lambda\mu\nu\kappa}$ we have denoted the Weyl tensor
for the metric $g$. One can easily show that the equation \ref{fwave}
is invariant with respect to the conformal rescaling i.e.
$f$ is a solution of $(\Box_g +\frac16 R)f=\frac12 Wf$ iff
$\tilde f=\Omega^3 f$ is a solution of
$({\Box}_{\tilde g} +\frac16 \tilde R)\tilde f=\frac12 \tilde W \tilde f$,
where all objects ${\Box}_{\tilde g}, \tilde R, \tilde W$ are calculated
with respect to the metric $\tilde g=\Omega^2 g$.

\section{Functionals $\QYK$ on initial spacelike surface}
\label{xyEH}
\subsection{Unconstrained degrees of freedom on $\Sigma_t$}
In \cite{kerrnut} it was shown that
 a pair of solutions $({\bf x}, {\bf y})$ of
the wave equation\footnote{The ``mono-dipole'' part of the field
has a special form related to the Poincar\'e charges
cf. \cite{JJspin2} and \cite{kerrnut}.}
gives a Weyl field in the following 2+2-form:
\[ W_{abcd}=-\frac12\rho^2{\bf x} \varepsilon_{ab}\varepsilon_{cd} \]
\[ W_{ABcd}=-\frac12\rho^2{\bf y} \varepsilon_{AB}\varepsilon_{cd} \]
\[ W_a{^B}{_{cd||B}} =-\frac12 \varepsilon_{cd} \varepsilon^b{_a}
\rho^3(\rho^{-1}{\bf x})_{,b} \] \be \label{weyl}
W{_{aBcd||E}}\varepsilon^{BE} =-\frac12 \varepsilon_{cd}
\rho^3(\rho^{-1}{\bf y})_{,a} \ee
\[ \kolo{W}_c{^{AB}}{_{d||AB}}=\frac14 \rho^4 {\bf x}_{cd}=\frac12
\rho^4 \left[ (\rho^{-2}{\bf x})_{,cd} -\frac12\eta_{cd} (\rho^{-2}{\bf
x})^{,b}{_{b}}\right] \]
\[ \kolo{W}_c{^{A}}{_{Bd||AC}} \varepsilon^{BC}=\frac14
\rho^4 \left[ \varepsilon_c{^b}(\rho^{-2}{\bf y})_{,bd} +
\varepsilon_d{^b}(\rho^{-2}{\bf y})_{,bc}  \right]
\]
where indices $A,B,C$ are related to the coordinates
$(\theta,\varphi)$
along spheres \[ S(r):=\{ t=\mbox{const}\, , \; r=\mbox{const}\} \]
 but the indices
 $a,b,c$  correspond to normal to $S(r)$ null coordinates
 $u=t-r$, $v=t+r$.
\subsection{Constraints for $E$ and $H$}
We have the following non-vanishing Christoffel symbols
for the three-metric $\eta_{kl}$ on $\Sigma_t$ in spherical
coordinates
$\Gamma^3{_{AB}}=-\frac1r\eta_{AB}$,
$\Gamma^A{_{3B}}= \frac1r\delta^A{_B}$, and
the spherical part $\Gamma^C{_{AB}}$
which is not dependent on radial coordinate $x^3:=r$
and defines covariant derivative on a sphere $r=$const which
we denote by ``$||$''. The angular coordinates $x^A$ parameterize
spheres $S(r)$. Let us also denote by $\dtwo$ a Laplace--Beltrami
operator on a unit sphere $S(1)$.

The 2+1-splitting of the constraint \ref{divEH}:
\be\label{E33} \partial_3(r^3E^{33}) + r^3 E^{3A}{_{||A}}=0 \, ,\ee
\be\label{Epolar} \partial_3(r^4E^{3A}{_{||A}}) + r^4\kolo{E}^{AB}{_{||AB}}
    -\frac12 r^2\dtwo E^{33} = 0 \, ,\ee
\be\label{Eaxial} \partial_3(r^4E^{3}{_{A||B}}\varepsilon^{AB})
    + r^4\kolo{E}_A{^B}{_{||BC}}   \varepsilon^{AC} =0 \ee
allows us to express explicitly all electric components of
the Weyl tensor in terms of $\bf x$ and $\partial_0{\bf y}$:
\be\label{E33x}
2 r^2E^{33} = {\bf x} \, , \quad 
2 r^2E_{3A||B}\varepsilon^{AB} = -\partial_0{\bf y} \ee
\be\label{E3AA}
2r^3 E^{3A}{_{||A}} = -2 \partial_3(r^3E^{33})=
-\partial_3(r{\bf x}) \ee
\be\label{EABpolar}
 2r^4\kolo{E}^{AB}{_{||AB}} = - 2\partial_3(r^4E^{3A}{_{||A}})
    +  \dtwo (r^2 E^{33}) = \partial_3\bigl(r\partial_3(r{\bf x}) \bigr) +
    \frac12 \dtwo {\bf x}
     \ee
\be\label{EABaxial}
     2r^4\kolo{E}_A{^B}{_{||BC}}   \varepsilon^{AC} =
     -2\partial_3(r^4E^{3}{_{A||B}}\varepsilon^{AB}) =
      \partial_3(r^2 \partial_0{\bf y})\ee
Similarly, we get the magnetic part
in terms of $\bf y$ and $\partial_0{\bf x}$:
\be\label{H33x}
2 r^2H^{33} = {\bf y} \ee
\be\label{H3ABy}
2 r^2H_{3A||B}\varepsilon^{AB} = -\partial_0{\bf x} \ee
\be\label{H3AA}
2r^3 H^{3A}{_{||A}} = 
-\partial_3(r{\bf y}) \ee
\be\label{HABpolar}
 2r^4\kolo{H}^{AB}{_{||AB}} = 
    \partial_3\bigl(r\partial_3(r{\bf y}) \bigr) +
    \frac12 \dtwo {\bf y}
     \ee
\be\label{HABaxial}
     2r^4\kolo{H}_A{^B}{_{||BC}}   \varepsilon^{AC} =
      \partial_3(r^2 \partial_0{\bf x})\ee

\subsection{Reduction of the quadratic forms and the functionals}
The ``spherical'' method from Appendix B in \cite{JJschwarzl}
can be easily applied for the reduction of the functional
$\int_V r^2(E^2 +H^2)$.
Let us for simplicity restrict ourselves to the case
of a three-dimensional ball $B(0,R)$ with radius $R$,
$V=B(0,R)=S^2\times [0,R]$, $\partial V=S(R)$,
$\int_V=\int_0^R \rd r \int_{S(r)}\rd\theta\rd\varphi$.
For exterior problems we also consider $V=\Sigma\setminus B(0,R)$.

The (2+1)-splitting of the tensor $q_{kl}$ gives the following
components on a sphere:\\
two scalars on $S^2$ -- $\stackrel{2}{q}:=\eta^{AB}q_{AB}$ and $q_{33}$,
vector $q_{3A}$ on $S^2$ and symmetric traceless tensor
$\stackrel{\circ}{q}\! {_{AB}}:=
q_{AB}-\frac12\eta_{AB}\stackrel{2}{q}$.
  Let $p^{kl}:= \sqrt{\det\eta_{ij}}q^{kl}$ be a tensor density on
$\Sigma$.
On each sphere $S(r)$ we can manipulate as follows
 \[
  \int_{S(r)}{p}^{kl}{q}_{kl} = \int_{S(r)} {p}^{33}{q}_{33} +
 2{p}^{3A}{q}_{3A}
+ \frac{1}{2} \stackrel{2}{p} \stackrel{2}{q} +
\stackrel{\circ}{p}\! {^{AB}}\stackrel{\circ}{q}\! {_{AB}}  =
\]
\[
=  \int_{S(r)} {p}^{33}{q}_{33} -
2(r{p}^{3A}{_{|| A}}) {\dtwo} ^{-1} (r{q}_{3A}{^{|| A}}) -
 2(r{p}^{3A|| B}\varepsilon_{AB}) {\dtwo} ^{-1}
(r{q}_{3A|| B}\varepsilon^{AB}) +
\frac{1}{2} \stackrel{2}{p} \stackrel{2}{q} +
\]
\[
+ 2 \int_{S(r)} (r^2\varepsilon^{AC}\stackrel{\circ}{p}\! {_A{^B}}{_{|| BC}})
 {\dtwo} ^{-1}({\dtwo}+2) ^{-1}
(r^2\varepsilon^{AC}\stackrel{\circ}{q}\! {_A{^B}}{_{|| BC}}) + \]
\[
+ 2 \int_{S(r)} (r^2\stackrel{\circ}{p}\! {^{AB}}{_{|| AB}})
 {\dtwo} ^{-1}({\dtwo}+2) ^{-1}
 (r^2\stackrel{\circ}{q}\! {^{AB}}{_{|| AB}}) \, ,
 \]
 where we have used the following identities on a sphere
 \be\label{dv}
  -\int_{S(r)} \pi^A v_A = (r{\pi}^{A}{_{|| A}}) {\dtwo} ^{-1}
(rv^{A}{_{|| A}})+ (r{\pi}^{A|| B}\varepsilon_{AB}) {\dtwo} ^{-1}
(rv_{A|| B}\varepsilon^{AB}) \, ,
 \ee
 and similarly for the traceless tensors we have
 \[
    \int_{S(r)} \stackrel{\circ}{\pi}\! ^{AB} \stackrel{\circ}{v}\! _{AB} =
  2 \int_{S(r)} (r^2\varepsilon^{AC}
  \stackrel{\circ}{\pi}\! {_A{^B}}{_{||BC}})
 {\dtwo} ^{-1}({\dtwo}+2) ^{-1}
(r^2\varepsilon^{AC}\stackrel{\circ}{v}\! {_A{^B}}{_{|| BC}}) + \]
 \be\label{ddtt}
+ 2 \int_{S(r)} (r^2\stackrel{\circ}{\pi}\! {^{AB}}{_{|| AB}})
 {\dtwo} ^{-1}({\dtwo}+2) ^{-1}
 (r^2\stackrel{\circ}{v}\! {^{AB}}{_{|| AB}}) \, .
 \ee
 The axial part of the quadratic form $\int_{S(r)}{p}^{kl}{q}_{kl}$
 we define as
 \[ \mbox{axial part}= -2\int_{S(r)}(r{p}^{3A|| B}\varepsilon_{AB})
{\dtwo} ^{-1} (r{q}_{3A|| B}\varepsilon^{AB}) +\]
 \[
+ 2 \int_{S(r)} (r^2\varepsilon^{AC}\stackrel{\circ}{p}\! {_A{^B}}{_{|| BC}})
 {\dtwo} ^{-1}({\dtwo}+2) ^{-1}
(r^2\varepsilon^{AC}\stackrel{\circ}{q}\! {_A{^B}}{_{|| BC}}) \, . \]
 The remainder we define as a polar part.
Using the above formulae by
inserting into them  $q_{kl}=H_{kl}$
and $q_{kl}=E_{kl}$ respectively we obtain the following expressions
\[ \hspace*{-0.5cm}
\mbox{axial part of} \int_{S(r)} r^2 H^{kl}H_{kl} =
  \frac12 \int_{S(r)} (r\dot{\bf x})(-\dtwo)^{-1}(r\dot{\bf x}) +
   \partial_r(r^2\dot{\bf x})\dtwo^{-1}(\dtwo +2)^{-1}
   \partial_r(r^2\dot{\bf x}) \]
\begin{eqnarray}
\mbox{polar part of} \int_{S(r)} r^2 E^{kl}E_{kl} & =&
  \frac12 \int_{S(r)} \frac32{\bf x}^2 +
  \partial_r(r{\bf x})(-\dtwo)^{-1}\partial_r(r{\bf x}) + \\
  \nonumber & & \hspace*{-2cm} 
   \bigl(
   \partial_r [ r\partial_r(r{\bf x})] +\frac12\dtwo{\bf x}\bigr)
   \dtwo^{-1}(\dtwo +2)^{-1}\bigl(
   \partial_r [ r\partial_r(r{\bf x})] +\frac12\dtwo{\bf x}\bigr)
    \end{eqnarray}
where we also used relations (\ref{E33x}--\ref{EABaxial}).
Similarly from (\ref{H33x}--\ref{HABaxial}) we obtain
\begin{eqnarray} & & \nonumber \hspace*{-1cm}
\mbox{polar} \int_{S(r)} r^2 E^{kl}E_{kl} +
\mbox{axial} \int_{S(r)} r^2 H^{kl}H_{kl} =
  \frac12 \int_{S(r)} \frac32{\bf y}^2 +
  \partial_r(r{\bf y})(-\dtwo)^{-1}\partial_r(r{\bf y}) + \\
  \nonumber & & + \frac12 \int_{S(r)}
   \bigl(
   \partial_r [ r\partial_r(r{\bf y})] +\frac12\dtwo{\bf y}\bigr)
   \dtwo^{-1}(\dtwo +2)^{-1}\bigl(
   \partial_r [ r\partial_r(r{\bf y})] +\frac12\dtwo{\bf y}\bigr)
   \\ & &
   + \frac12 \int_{S(r)} (r\dot{\bf y})(-\dtwo)^{-1}(r\dot{\bf y}) +
   \partial_r(r^2\dot{\bf y})\dtwo^{-1}(\dtwo +2)^{-1}
   \partial_r(r^2\dot{\bf y})
    \end{eqnarray}
This way we obtain \eq{BRV}.

The similar manipulations give the following
\begin{eqnarray}
 \int_{S(r)} r^4 E_{kr}E^{kr} & = & \nonumber
 \int_{S(r)} \Bigl[ (r^2 E_{rr})^2 -
 (r^3 E^{rA}{_{||A}})\dtwo^{-1} (r^3 E^{rA}{_{||A}}) \\
 &  & \nonumber
  - (r^3 E_{rA||B}\varepsilon^{AB})\dtwo^{-1}
  (r^3 E_{rA||B}\varepsilon^{AB}) \Bigr]
 \\ & = & \nonumber \frac14 \int_{S(r)}
  {\bf x}^2
 + \partial_r(r{\bf x})(-\dtwo)^{-1}\partial_r(r{\bf x})
 +
  (r\dot{\bf y})(-\dtwo)^{-1}(r\dot{\bf y})
 \end{eqnarray}
 and similarly
\begin{eqnarray}
 \int_{S(r)} r^4 H_{kr}H^{kr} & = & \nonumber
  \frac14 \int_{S(r)}
  {\bf y}^2
 + \partial_r(r{\bf y})(-\dtwo)^{-1}\partial_r(r{\bf y})
 +
  (r\dot{\bf x})(-\dtwo)^{-1}(r\dot{\bf x})
 \end{eqnarray}
 which finally gives \eq{EMV}.

 It is also instructive to see how the conservation law for
 the functional \eq{EMV} can be obtained straightforward from the
  wave equations
 \be\label{box} \Box {\bf x} =0 \, , \quad \Box {\bf y} =0 \, .\ee
 This can be shown as follows
\begin{eqnarray} \nonumber
  \partial_0 & & \hspace*{-6ex} \frac12 \int_V \left[ {\bf x}^2
 + \partial_r(r{\bf x})(-\dtwo)^{-1}\partial_r(r{\bf x})
 + (r\dot{\bf x})(-\dtwo)^{-1}(r\dot{\bf x})
 \right]\rd r \sin\theta\rd\theta\rd\varphi = \\ & =& \nonumber
  \int_V \left[ {\bf x}\dot{\bf x}
 + \partial_r(r{\bf x})(-\dtwo)^{-1}\partial_r(r\dot{\bf x})
 + (r\ddot{\bf x})(-\dtwo)^{-1}(r\dot{\bf x})
 \right]\rd r \sin\theta\rd\theta\rd\varphi =  \\ & =& \nonumber
  \int_V \left[ \dtwo {\bf x}
 +r \partial_r (\partial_r(r{\bf x}))
 - r^2\ddot{\bf x}
 \right] \dtwo^{-1}\dot{\bf x}\rd r \sin\theta\rd\theta\rd\varphi
 \\ & & \nonumber
 + \int_{\partial V} \partial_r(r{\bf x})(-\dtwo)^{-1}(r\dot{\bf x})
 \sin\theta\rd\theta\rd\varphi \\ & =& \nonumber
 \int_V
  r^2\Box{\bf x} \dtwo^{-1}\dot{\bf x}\rd r \sin\theta\rd\theta\rd\varphi
 + \int_{\partial V} \partial_r(r{\bf x})(-\dtwo)^{-1}(r\dot{\bf x})
 \sin\theta\rd\theta\rd\varphi
 \end{eqnarray}
 The volume term vanishes because of the wave equation \eq{box}
 and to get the result
 \be\label{d0} \partial_0 \int_V \left[ {\bf x}^2
 + \partial_r(r{\bf x})(-\dtwo)^{-1}\partial_r(r{\bf x})
 + (r\dot{\bf x})(-\dtwo)^{-1}(r\dot{\bf x})
 \right]\rd r \sin\theta\rd\theta\rd\varphi =0 \ee
 we need to assume that $\bf x$
 or rather $\dot{\bf x}$ vanishes on the boundary.

The other examples of our functionals $\QYK$
on surface $\Sigma_0$ can be expressed as follows:
\begin{eqnarray} \nonumber
\hspace*{-0.8cm}
\QYK({\cal T}_0,V;{\cal T}_i\wedge{\cal T}_0) & = &
\int_V \TEM\bigl({\cal T}_0,{\cal T}_0,
F(W,{\cal T}_i\wedge{\cal T}_0)\bigr)\rd^3 x  \\ \nonumber
&=& \frac12 \int_V \left(
 E_{ki}E^{k}{_i} + H_{ki}H^{k}{_i} \right)r^2\sin\theta\rd r
 \rd\theta\rd\varphi \\
 & =& \label{EMV0z} \frac12 \int_V \Bigl[
 (vE^{kr}+rv_{||A}E^{kA})(vE_{kr}+rv_{||A}E_k{^A})
 \\ & & \nonumber
 +(vH^{kr}+rv_{||A}H^{kA})(vH_{kr}+rv_{||A}H_k{^A})
 \Bigr]r^2\rd r \sin\theta\rd\theta\rd\varphi
 \\ & = & \nonumber \frac12 \int_V v^2\Bigl[
 E^{kl}E_{kl} +\frac12 r^2 (E^{kA}E_k{^B})_{||AB}
 - r (E^{kA}E_{kr})_{||A}
 \\ &  & \nonumber +
 H^{kl}H_{kl} +\frac12 r^2 (H^{kA}H_k{^B})_{||AB}
 - r (H^{kA}H_{kr})_{||A}
 \Bigr]r^2\rd r \sin\theta\rd\theta\rd\varphi \, ,
 \end{eqnarray} where $v=\frac{x_i}r$,
\begin{eqnarray} \nonumber
\hspace*{-0.7cm}
\QYK({\cal T}_0,V;{\cal T}_i\wedge{\cal D}) & = &
\int_V \TEM\bigl({\cal T}_0,{\cal T}_0,
F(W,{\cal T}_i\wedge{\cal D})\bigr)\rd^3 x  \\
 & =& \nonumber \frac12 \int_V v_{||A}v_{||C}
 \varepsilon^{AB}\varepsilon^{CD}\Bigl(E^{kB}E_k{^D}+H^{kB}H_k{^D}
 \Bigr)r^6\rd r \sin\theta\rd\theta\rd\varphi
 \\  & = &  \label{EMDz}
 \frac12 \int_V v^2\Bigl[
  E^{kA}E_{kA}+ H^{kA}H_{kA} \\ & & \nonumber \hspace*{0.5cm}
 -\frac12 r^2 \varepsilon^{AB}\varepsilon^{CD}
 (E_{kB}E^k{_D}+ H_{kB}H^k{_D})_{||AC}
 \Bigr]r^4\rd r \sin\theta\rd\theta\rd\varphi \, ,
 \end{eqnarray}
and for $Q_k^{{\scriptscriptstyle\rm boost}}:= {\cal D}\wedge{\cal L}_{0k}
-\frac12x^\mu x_\mu{\cal T}_0\wedge{\cal T}_k$ we get
\begin{eqnarray} \nonumber
\hspace*{-0.7cm}
\QYK({\cal T}_0,V;Q_i^{{\scriptscriptstyle\rm boost}}) & = &
\int_V \TEM\bigl({\cal T}_0,{\cal T}_0,
F(W,Q_i^{{\scriptscriptstyle\rm boost}})\bigr)\rd^3 x  \\ 
 & =& \label{EML0z} \frac12 \int_V r^4\Bigl[
 (vE^{kr}-rv_{||A}E^{kA})(vE_{kr}-rv_{||B}E_k{^B})
 \\ & & \nonumber
 +(vH^{kr}-rv_{||A}H^{kA})(vH_{kr}-rv_{||B}H_k{^B})
 \Bigr]r^2\rd r \sin\theta\rd\theta\rd\varphi
 \\ & = & \nonumber \frac12 \int_V v^2\Bigl[
 E^{kl}E_{kl} +\frac12 r^2 (E^{kA}E_k{^B})_{||AB}
 + r (E^{kA}E_{kr})_{||A}
 \\ &  & \nonumber +
 H^{kl}H_{kl} +\frac12 r^2 (H^{kA}H_k{^B})_{||AB}
 + r (H^{kA}H_{kr})_{||A}
 \Bigr]r^6\rd r \sin\theta\rd\theta\rd\varphi \, .
 \end{eqnarray}
Hence, from the identity
$\displaystyle\sum_{i=1}^3 \left(\frac{x_i}{r}\right)^2=1$ \
 one can easily check that
\be\label{BR4T}
 2\sum_{k=1}^3 \int_V \TEM\bigl({\cal T}_0,{\cal T}_0,
F(W,{\cal T}_k\wedge{\cal T}_0)\bigr)\rd^3 x =
\int_V
 \TBR\bigl({\cal T}_0,{\cal T}_0,{\cal T}_0,{\cal T}_0
\bigr)\rd^3 x \ee
\be\label{BRK3T}
 2\sum_{k=1}^3 \int_V \TEM\bigl({\cal K}_0,{\cal T}_0,
F(W,{\cal T}_k\wedge{\cal T}_0)\bigr)\rd^3 x =
\int_V
 \TBR\bigl({\cal K}_0,{\cal T}_0,{\cal T}_0,{\cal T}_0
\bigr)\rd^3 x \ee
\be\label{BRK3Tbis}
 2\sum_{\mu=0}^3 \int_V \TEM\bigl({\cal T}_0,{\cal T}_0,
F(W,{\cal T}_\mu\wedge{\cal D})\bigr)\rd^3 x =
\int_V
 \TBR\bigl({\cal K}_0,{\cal T}_0,{\cal T}_0,{\cal T}_0
\bigr)\rd^3 x \ee
\be\label{BR2K2Tbis}
 2\sum_{\mu=0}^3 \int_V \TEM\bigl({\cal K}_0,{\cal T}_0,
F(W,{\cal T}_\mu\wedge{\cal D})\bigr)\rd^3 x =
\int_V
 \TBR\bigl({\cal K}_0,{\cal K}_0,{\cal T}_0,{\cal T}_0
\bigr)\rd^3 x \ee
\be  \label{BR2K2T}
2\sum_{k=1}^3 \int_V \TEM\bigl({\cal T}_0,{\cal T}_0,
F(W,Q_k^{{\scriptscriptstyle\rm boost}})\bigr)\rd^3 x
= \int_V
 \TBR\bigl({\cal K}_0,{\cal K}_0,{\cal T}_0,{\cal T}_0
\bigr)\rd^3 x \, .
 \ee
\be  \label{BR3K1T}
2\sum_{k=1}^3 \int_V \TEM\bigl({\cal K}_0,{\cal T}_0,
F(W,Q_k^{{\scriptscriptstyle\rm boost}})\bigr)\rd^3 x
= \int_V
 \TBR\bigl({\cal K}_0,{\cal K}_0,{\cal K}_0,{\cal T}_0
\bigr)\rd^3 x \, .
 \ee
The above formulae show explicitly how the functionals
$\QYK$ include the functionals $\QBR$.

\section{Index of symbols}
\begin{itemize}
\item[$M$] --  spacetime 
\item[$;\mu$] -- four-dim. covariant derivative also denoted by $\nabla_\mu$
\item[$\varepsilon_{\mu \nu \gamma\delta}$] -- Levi--Civita
skew-symmetric tensor in spacetime $M$
\item[$*$] -- Hodge dual operation,
$ t^*_{\mu\nu} := \frac 12 \varepsilon_{\mu\nu}{^{\lambda\kappa}}
t_{\lambda \kappa}$
\item[$\partial_{\mu}$] -- partial derivative also denoted by "," (comma)
\item[$\eta_{\mu\nu} $] -- Minkowski metric
\item[$x^\mu$] -- Cartesian coordinates in Minkowski spacetime
\item[${\cal T}_{\mu}$] -- translation Killing field,
${\cal T}_{\mu}:=\base x\mu$
\item[${\cal L}_{\mu\nu}$] -- boost or rotation Killing field,
 ${\cal L}_{\mu\nu}:=x_{\mu} \base x\nu - x_\nu \base x\mu$
\item[${\cal D}$] -- scaling conformal Killing field,
 ${\cal D}:=x^\nu \base x\nu$
\item[${\cal K}_{\mu}$] -- ``acceleration'' conformal Killing field,
${\cal K}_\mu := -2x_\mu{\cal D} + x^\sigma x_\sigma {\cal T}_\mu$
\item[$t$] --  time coordinate 
\item[$r$] --  radial coordinate 
\item[$x$] --  ``inverse'' radial coordinate $r=x^{-1}$
\item[$u,v$] -- null coordinates on $M$, $u=t-r$, $v=t+r$
\item[$\theta,\phi $] -- spherical coordinates on $S^2$
\item[$\rd $] -- exterior derivative
\item[$\mu,\nu,\ldots $] -- four-dimensional indices running $0,\ldots,3$
\item[$k,l,\ldots $] -- three-dimensional indices running $1,\ldots,3$
\item[$A,B,\ldots $] -- two-dimensional indices on a sphere
\item[$\Box$] -- d'Alambertian, wave operator
\item[$R$] -- scalar curvature
\item[$X$] -- vector field
\item[$i^0$] -- spatial infinity
\item[$\scri$] -- null infinity
\item[$\scri^+$] -- future null infinity
\item[$\scri^-$] -- past null infinity
\item[$\delta^\mu{_\nu} $] -- Kronecker's delta
\item[$S(r)$] -- sphere with radius $r$ parameterized by $\theta,\phi$
\item[$\eta_{AB}$] -- metric on $S(r)$
\item[$\kolo{t}_{AB}$] -- two-dim. traceless part of the tensor ${t}_{AB}$,
$\stackrel{\circ}{t}_{AB} := t_{AB} - \frac12 \eta_{AB}
\eta^{CD}t_{CD}$
\item[$||$] -- two-dimensional covariant derivative on a sphere
\item[$\dtwo$] -- two-dimensional laplacian on a unit sphere
\item[$\varepsilon^{AB}$] -- two-dimensional skew-symmetric tensor on $S(r)$,
$r^2\sin\theta\varepsilon^{\theta\phi}=1$
\item[$\varepsilon^{ab}$] -- skew-symmetric tensor on the
two-dimensional space orthogonal to $S(r)$,
$\varepsilon^{uv}=2$
\item[$R^\mu{_{\nu\lambda\sigma}} $] -- curvature tensor
\item[$R_{\mu\nu}$] -- Ricci tensor
\item[$\Gamma^\lambda{_{\mu\nu}} $] -- Christoffel symbol
\item[$W{_{\mu\nu\lambda\sigma}} $] -- spin-2 field,  Weyl tensor
\item[${\bf x},{\bf y}$] -- gauge-invariants
\item[${p}_\mu$] -- four-momentum charge
\item[${b}_\mu$] -- dual four-momentum charge
\item[${j}_{\mu\nu}$] -- angular momentum and static moment charges
\item[${w}_{\mu\nu}$] -- ``ofam'' charge
\item[$Q_{\mu\nu}$] -- CYK tensor
\item[${\cal Q}_{\mu\nu\lambda}$] -- CYK equation
\item[$\Omega$] -- conformal factor
\item[$F_{\mu\nu}$] -- Maxwell field
\item[$\TEM_{\mu\nu}$] -- energy-momentum tensor for Maxwell field
\item[$\TBR_{\mu\nu\alpha\beta}$] -- Bel--Robinson tensor
\item[$\QEM$] -- conserved quantity for Maxwell field
\item[$\QBR$] -- conserved quantity for Bel--Robinson tensor
\item[$\QYK$] -- conserved quantity for CYK tensor bilinear in Weyl field
\item[$\Sigma$] -- spacelike hypersurface
\item[$E$] -- electric part of Weyl field
\item[$H$] -- magnetic part of Weyl field
\item[$T_{\mu\nu\alpha\beta\gamma\delta}$] -- (super-)tensor

\end{itemize}


\begin{thebibliography}{666}

\bibitem{Baleanu} D. Baleanu, Nuovo Cimento B 114, 1065--1072 (1999)

\bibitem{BCK} I.M. Benn, P. Charlton and J. Kress, Journal of
 Mathematical Physics 38, 4504--4527 (1997)

\bibitem{BF} S. Benenti, M. Francaviglia, General Relativity and Gravitation,
Edited by A. Held, vol. 1, 393--439, Plenum Press, (New York 1980)

\bibitem{BB} I. Bia{\l}ynicki--Birula and Z. Bia{\l}ynicka--Birula,
{\it Quantum Electrodynamics}, Pergamon, (Oxford 1975)

\bibitem{Ch-Kl} D. Christodoulou, S. Klainerman, Communications on Pure
and Applied Mathematics 43,  137--199 (1990)

\bibitem{Ch-Kl1} D. Christodoulou and S. Klainerman,
{\it The Global Nonlinear Stability of the Minkowski Space},
 Princeton University Press (1993)

\bibitem{ptced} P.T. Chru\'sciel and E. Delay,
 {\it Existence of non-trivial, vacuum, asymptotically simple
 space-times}, gr-qc/0203053

\bibitem{ptcjjk} P.T. Chru\'sciel, J. Jezierski and J. Kijowski,
{\it Hamiltonian Field Theory in the Radiating Regime},
 Lecture Notes in Physics: m70, Springer (2002)

\bibitem{Collinson} C.D. Collinson and L. Howarth, General Relativity and
Gravitation 32, 1767--1776 (2000); General Relativity and
Gravitation 32, 1845--1849 (2000)

\bibitem{CS} J. Corvino, Commun. Math. Phys. 214, 137--189 (2000);
J. Corvino and R. Schoen, {\it Vacuum spacetimes which are identically
Schwarzschild near spatial infinity}, talk given at the Santa Barbara
Conference on Strong Gravitational Fields, June 22-26 (1999),
\verb+ http://doug-pc.itp.ucsb.edu/online/gravity_c99/schoen/ +

\bibitem{Dietz-Rudiger} W. Dietz and R. Rudiger, Proc. Roy. Soc. Lond.
 A Mat. 37, 361--378 (1981)

\bibitem{HF} H. Friedrich, {\it Asymptotic Structure of Space-Time},
in: Recent Advances in General Relativity, Edited by A.I. Janis
 and J.R. Porter, Birkh\"auser, (Basel 1992)


\bibitem{TF} T. Friedrich, {\it Self--duality of Riemannian Manifolds
and Connections}, in: Self--dual Riemannian Geometry and Instantons,
TEUBNER--TEXTE zur Mathematik, Band 34, (Leipzig 1981)

\bibitem{GR} G.W. Gibbons, P.J. Ruback, Commun. Math. Phys. 115, 267--300
(1988)

\bibitem{JNG} J.N. Goldberg, Physical Review D 41, 410--416 (1990)

\bibitem{Ibohal} N.G. Ibohal, Astrophysics and Space Science 249,
 73--93 (1997)



\bibitem{JJspin2} J. Jezierski, General Relativity and
Gravitation 27, 821--843 (1995)

\bibitem{kerrnut} J. Jezierski,
 Classical and Quantum Gravity 14, 1679--1688 (1997)

\bibitem{JJschwarzl} J. Jezierski, {\it Gauge-invariant formulation of the
linearized Einstein equations  on the Schwarzschild background}, in
Current Topics in Math. Cosmology, eds M.Rainer and H-J Schmidt, World
Scientific (1998); 
General Relativity and Gravitation 31,   1855-1890 (1999),
(gr-qc/9801068)

 \bibitem{JJnullweyl} J. Jezierski,
  Classical and Quantum Gravity 19,   2463--2490 (2002)

\bibitem{Pen-Rin} R. Penrose and W. Rindler, {\it Spinors and Space-time},
Cambridge University Press, Vol. 2, p.396 (Cambridge 1986)

\bibitem{Rietdijk} R.H. Rietdijk, Classical and Quantum Gravity {7},
 247 (1990)

\bibitem{Sen} J.M.M. Senovilla, 
 Classical and Quantum Gravity 17,  2799--2842 (2000) 

\bibitem{tafel} J. Tafel and S. Pukas, Classical and Quantum Gravity 17,
 1559--1570 (2000); J. Tafel, Classical and Quantum Gravity 17,
 4397--4408 (2000)

\bibitem{vanHolten} J.W. van Holten, Phys. Lett. B {342}, 47 (1995)

\bibitem{Visinescu} D. Vaman and M. Visinescu, Phys. Rev. D {54},
 1398--1402 (1996)

\bibitem{Yano} K. Yano, Ann. Math. 55, 328 (1952)


\end{thebibliography}
\end{document}